\documentclass[%
 reprint,
 amsmath,amssymb,
 aps,
 floatfix,
]{revtex4-2}

\usepackage{graphicx}
\usepackage{dcolumn}
\usepackage{bm}
\usepackage{hyperref}
\hypersetup{
    colorlinks=true,
    linkcolor=blue,
    filecolor=blue,      
    urlcolor=blue,
    citecolor=blue,
}
\usepackage{url}
\usepackage[mathlines]{lineno}

\begin{document}

\title{Accelerated protons produced by magnetic Penrose process in Sgr A*}

\author{Myeonghwan Oh}
 \email{projectwrist@gmail.com}
\affiliation{%
  Department of Astronomy and Atmospheric Sciences, Kyungpook National University, Daegu, 41566, Republic of Korea
}%
\author{Myeong-Gu Park}%
 \email{mgp@knu.ac.kr}
\affiliation{%
  Department of Astronomy and Atmospheric Sciences, Kyungpook National University, Daegu, 41566, Republic of Korea
}%

\date{\today}

\begin{abstract}
Typical mechanisms to extract energies from a rotating black hole are the Blandford-Znajek process and the Penrose process. The Penrose process requires a special condition that is difficult to occur in common astrophysical situations. However, the magnetic Penrose process (MPP) does not require such a special condition, and can produce ultrahigh energy cosmic rays. When neutrons decay near a rotating black hole, the MPP efficiency of the produced proton is maximized. The supermassive black hole in Sagittarius A* (Sgr A*) is likely to have a radiatively inefficient accretion flow that is hot enough to produce neutrons by nuclear reactions, which can be subsequently accelerated to high-energy by the MPP. We calculate the production rate of accelerated protons from the Sgr A* to estimated the gamma ray flux at Earth produced by these accelerated protons and the flux of the accelerated protons themselves transported from Sgr A* to Earth. We find that these very high-energy gamma rays ($E_{\gamma}\gtrsim10\,\mathrm{TeV}$) amount to a significant fraction of the flux of the gamma ray from the HESS J1745-290 and the central molecular zone around $100\,\mathrm{TeV}$. The accelerated proton flux, when the dimensionless spin parameter $a_{*}=0.5$ and the magnetic field strength in the vicinity of the black hole $B_{0}=100\,\mathrm{G}$, is about $1.6-4.1\%$ of the cosmic ray proton flux from KASCADE experiment at about $1\,\mathrm{PeV}$. Due to the finite decay time of neutrons that need to be transported from the accretion flow to the acceleration zone, our acceleration model can operate only around black holes with mass not much greater than $\sim10^8\,M_\odot$.
\end{abstract}

\maketitle


\section{Introduction}

After the development of the black hole thermodynamics and the discovery of the irreducible mass of a rotating black hole \cite{christodoulou1970reversible,christodoulou1971reversible,hawking1971gravitational}, many mechanisms of extraction of the energy from a rotating black hole (Kerr black hole) have been presented. The Penrose process, one of such mechanisms, extracts the energy by particle decay \cite{penrose1969gravitational}. The observer at infinity sees one of the two fragments fall into the black hole with a negative energy and also the other fragment escapes to infinity with a positive energy. Through this process, the black hole loses angular momentum and the rotational energy. However, the Penrose process requires the relative velocity between the two fragments to be larger than half the speed of light. Therefore, the normal Penrose process is not generally expected in a typical astrophysical jet engine \cite{bardeen1972rotating,wald1974energy}.

About a decade later, Wagh, Dhurandhar, and Dadhich \cite{wagh1985revival} proposed that energy from electromagnetic potential can alleviate the constraint on the relative velocity. When particle 1 is the incident neutral particle and particle 2 one of two charged fragments,
\begin{equation}
\label{WBineq}
    \gamma\left(E_{1}-\left|\mathbf{v}\right|\sqrt{{E_{1}}^{2}+g_{00}}\right)-e_{2}\phi\leq0,
\end{equation}
where $E_{1}$ is the rest mass energy of particle 1, $\mathbf{v}$ the relative velocity in the rest frame of particle 1, $\gamma$ the Lorentz factor from the relative velocity $\mathbf{v}$, $\phi$ an electric potential and $e_{2}$ an electric charge of particle 2. Equation \eqref{WBineq} is the Wald-Bardeen inequality with an electromagnetic interaction and presents the minimum condition of an energy extraction from the magnetic Penrose process (hereafter MPP). When the relative velocity $\mathbf{v}=0$, the Eq. \eqref{WBineq} becomes $E_{1}\leq\left|e_{2}\phi\right|$. Therefore, even when $\mathbf{v}=0$, through the condition of $E_{1}\leq\left|e_{2}\phi\right|$ energy can be extracted through the MPP. Also, the inequalities showed that the MPP is possible without any constraints on the relative velocity.

The efficiency of the Penrose process is defined as
\begin{equation}
    \eta\equiv\frac{E_{3}-E_{1}}{E_{1}},
\end{equation}
where $E_{1}$ is the incident particle energy and $E_{3}$ is the escape particle energy. The maximum efficiency of the Penrose process is $20.7\%$, while Parthasarathy $\mathit{et}$ $\mathit{al.}$ presented higher efficiency $\eta>1$ with the MPP \cite{parthasarathy1986high}. Afterwards, Narayan, McClintock, and Tchekhovskoy showed that the efficiency of the MPP can reach to few hundred percent with general relativistic magnetohydrodynamics (GRMHD) calculations \cite{narayan2014energy}.

The MPP does not have a constraint on the relative velocity of fragments, and nor a constraint on the magnetic field strength around the black hole. Blandford-Znajek process has a threshold magnetic field strength to produce electron-positron pair cascade for force-free condition \cite{blandford1977electromagnetic} with
\begin{equation}
\label{thresB}
    B\approx6.2\times10^{4}\left(\frac{M}{a}\right)^{\frac{3}{4}}\left(\frac{10\,M_{\odot}}{M}\right)^{\frac{1}{2}}\,\mathrm{G}.
\end{equation}
For a $10\,M_{\odot}$ black hole and the Sagittarius A* (Sgr A*), the supermassive black hole in the center of the Galaxy with mass of about $4\times10^{6}\,M_{\odot}$ \cite{abuter2019geometric}, such threshold magnetic fields are about $10^{4}\,\mathrm{G}$ and about few $10^{2}\,\mathrm{G}$, respectively. But, in the MPP, the magnetic field strength required is only a few $\mathrm{mG}$ and a few $10^{-8}\,\mathrm{mG}$ to reach $100\%$ efficiency for $10\,M_{\odot}$ black hole and the Sgr A*, respectively \cite{dadhich2018distinguishing}.

To analyze the motion of diverged charged particles from MPP, one should know the motion of a charged particle in the electromagnetic field near a black hole. Since accretion flow tends to follow a Kerr black hole for axial symmetry, one can use the Wald solution which is the solution for the Maxwell equation when a Kerr black hole is immersed in an aligned test uniform magnetic field to describe the electromagnetic field near the astrophysical black hole for leading order \cite{wald1974black, stuchlik2016acceleration,tursunov2016circular,kolovs2017possible,ruffini2019gev,rueda2020blackholic,tursunov2020supermassive,moradi2021newborn,rueda2022gravitomagnetic,rastegarnia2022structure}. When such an electromagnetic field and special condition is met, charged particles can escape to infinity with ultrahigh energy \cite{stuchlik2016acceleration,tursunov2020supermassive}.

Tursunov $\mathit{et}$ $\mathit{al.}$ (2020) suggested that charged particles produced through neutron decay can be accelerated to ultrahigh energy through MPP and calculated trajectories of the escaping particle in the vicinity of SMBHs like Sgr A* and M87* \cite{tursunov2020supermassive}. They showed that, unlike the Penrose process that is limited in the ergosphere, MPP can occur in a broad region, if black hole mass and uniform magnetic field strength are large enough. They also showed that the energy of protons accelerated by the MPP around Sgr A* can reach $\simeq10^{15.6}\,\mathrm{eV}$, which corresponds to the knee of cosmic rays spectrum, and showed the compatibility with the HESS collaboration analysis. The HESS collaboration argues that from the gamma-ray data, petaelectronvolt protons are produced at the Galactic Center (GC), and they estimated that the Sgr A* is the pevatron \cite{collaboration2016acceleration}.

However, Tursunov $\mathit{et}$ $\mathit{al.}$ (2020) did not concretely present the origin of the neutrons and did not estimate the flux of the ultrahigh energy proton at Earth. An accretion disk that is hot enough to produce neutrons from thermonuclear reactions can easily become the source of neutrons \cite{aharonian1984gamma,jean2001neutron,kafexhiu2019nuclear}.

In 1976, Shapiro, Lightman, and Eardley \cite{1976ApJ...204..187S} presented the high ion temperature accretion flow that is sufficiently hot for nuclear reactions to produce neutrons. However, the model is thermally unstable \cite{park1995stability,narayan1995advection}. Narayan and Yi \cite{narayan1994advection,narayan1995advection} presented a thermally stable advection-dominated accretion flow (ADAF) model in which ions are kept at a very high temperature. Reference~\cite{narayan1994advection} found self-similar solutions of such accretion flow that show the accreting gas temperature is close to the virial temperature and the disk structure is quasispherical. In the ADAF model, due to the large mass difference between ions and electrons, ions dominate the heating rate $q^{+}$ while electrons dominate the cooling rate $q^{-}$. Thus, the ion temperature is about $10^{3}$ times higher than the electron temperature. The ADAF was employed to explain many x-ray sources, including Sgr A* \cite{narayan1997advection,narayan1998advection}.

The accretion flow into the Sgr A* is considered to be a radiatively inefficient accretion flow (RIAF), because the Sgr A* has a low luminosity. Therefore, the accretion disk of the Sgr A* will be sufficiently hot to produce neutrons from nuclear reactions. Also, ADAF parameters and the accretion rate of the Sgr A* are reasonably well known. Thus, one can estimate the neutron number that flows into the magnetic field near the Sgr A*.

The estimation of HESS collaboration, the high-energy efficiency of the MPP, and the accretion flow of Sgr A* show that the MPP acceleration model with neutron decay is suitable for Sgr A*.

The MPP protons have a maximum energy of about $10^{15.6}\,\mathrm{eV}$. So the gyroradius of the proton does not exceed the scale of the random magnetic field in the Galaxy \cite{ptuskin2006cosmic}. Thus, the MPP protons will be confined and diffuse out in the Galaxy by scattering. Therefore, if we calculate the production rate of the accelerated protons from the Sgr A* through the MPP, we can estimate how much the accelerated protons from the Sgr A* will affect the cosmic ray knee observed at Earth.

We conduct a study on how many protons that are accelerated in Sgr A* by the MPP model of Tursunov $\mathit{et}$ $\mathit{al.}$ (2020) impact the spectra of gamma rays from the GC and the cosmic ray spectrum. In our model, first, neutrons are produced in the accretion flow of the black hole by nuclear reactions. Subsequently, the neutrons flow into the magnetic field of the vicinity of the black hole (Sec. \uppercase\expandafter{\romannumeral2}). The neutrons that flow near the black hole decay into protons and electrons. The electrons fall into the black hole and the protons are accelerated by the MPP and escape to infinity from the black hole (Sec. \uppercase\expandafter{\romannumeral3}). Finally, escaped protons diffuse out within the Galaxy and produce gamma rays in the GC (Sec. \uppercase\expandafter{\romannumeral4}) and cosmic rays flux (Sec. \uppercase\expandafter{\romannumeral5}).

\section{Neutron production for ADAF}
We adopt the hot ADAF model to describe the accretion flow in Sgr A*. Narayan and Yi \cite{narayan1994advection,narayan1995advection} presented self-similar solutions under Newtonian gravity. Radial velocity $v_r$ \eqref{ADAF1}, rotation angular velocity $\Omega$ \eqref{ADAF2}, mass density $\rho$ \eqref{ADAF3}, and disk scale height $H$ \eqref{ADAF5} are described by simple power laws in radius: 
\begin{eqnarray}
\label{ADAF1}
v_{r} &=& -2.12 \times 10^{10}\alpha c_{1}r^{-1/2}\,\mathrm{cm}\,\mathrm{s^{-1}},\\
\label{ADAF2}
\Omega &=& 7.19 \times 10^{4} c_{2}m^{-1}r^{3/2}\,\mathrm{s^{-1}},\\
\label{ADAF3}
\rho &=& 3.79 \times 10^{-5}\alpha^{-1} c^{-1}_{1}c^{-1/2}_{3}m^{-1}\dot{m}r^{-3/2}\,\mathrm{g\,cm^{-3}},\\
\label{ADAF5}
H &=& (2.5c_{3})^{1/2}R,
\end{eqnarray}
where $r$, $m$, and $\dot{m}$ are scaled disk radius, black hole mass, and mass accretion rate with $r\equiv R/R_{s}$, $m\equiv M/M_{\odot}$ and $\dot{m}\equiv\dot{M}/\dot{M}_{\mathrm{Edd}}$, respectively. The Schwarzschild radius $R_{s}\equiv 2GM/c^{2}$ and the Eddington rate $\dot{M}_{\mathrm{Edd}}\equiv L_{\mathrm{Edd}}/(0.1c^{2})$ where $L_{\mathrm{Edd}}$ is the Eddington luminosity. 
Ion temperature $T_i$ and electron temperature $T_e$ are also described by simple power law in radius:
\begin{equation}
\label{ADAF4}
T_{i}+1.08T_{e} = 6.66 \times 10^{12}\beta c_{3}r^{-1}\,\mathrm{K}.
\end{equation}
The solutions depend on the parameters $\alpha$, $\beta$, and $f$, which describe the physical processes in ADAF: $\alpha$ is the viscosity parameter, $\beta$ the fraction of the gas pressure to the total pressure, and $f$ the fraction of viscously dissipated energy which is advected. Constants $c_{1}$ and $c_{3}$ are defined by parameters $\alpha$, $\beta$, and $f$ \cite{narayan1994advection} with
\begin{eqnarray}
c_{1}\equiv\frac{\left(5+2\epsilon'\right)}{3\alpha^{2}}g\left(\alpha,\epsilon'\right),\quad c_{3}\equiv\frac{2\left(5+2\epsilon'\right)}{9\alpha^{2}}g\left(\alpha,\epsilon'\right),
\end{eqnarray}
where 
\begin{eqnarray}
g\left(\alpha,\epsilon'\right)&\equiv&\left[1+\frac{18\alpha^{2}}{\left(5+2\epsilon'\right)^{2}}\right]^{1/2}-1,\nonumber\\
\epsilon'&\equiv&\frac{\epsilon}{f}=\frac{1}{f}\left(\frac{5/3-\gamma}{\gamma-1}\right),\nonumber\\
\gamma&=&\frac{32-24\beta-3\beta^{2}}{24-21\beta}
\end{eqnarray}

Narayan, Kato, and Honma (hereafter NKH) used the Paczyñsky-Wiita potential \cite{paczynsky1980thick} that mimics the relativistic effects of a Schwarzschild black hole \cite{narayan1997global}. Gammie and Popham (hearafter GP) presented fully relativistic ADAF solutions in the Kerr geometry \cite{gammie1998advection,popham1998advection}. These models are more relevant to our scenario. However, one has to solve the two-point boundary value problems with singularity while regulating a sonic point in the NKH model, and also one has to solve the boundary value problems for a sonic point and viscous point in the GP model whereas one can get analytic solutions from self-similar solutions. In addition, NKH and GP models does not provide information about the vertical structure of the flow, because these models are 1D solutions, and these ADAF models do not reflect the complex magnetic field structure around the black hole nor the inner region of the accretion flow, which will be undoubtedly complex. Therefore, we adopt the self-similar solutions that should be good enough approximations for our study, considering all the uncertainties.

We choose ADAF parameters from previous studies based on observation. First, we select the accretion rate. The accretion rate of the Sgr A* has been estimated by Faraday rotation \cite{agol2000sagittarius,marrone2006unambiguous,sharma2007faraday,shcherbakov2012sagittarius,bower2018alma}. References~\cite{sharma2007faraday,shcherbakov2012sagittarius} estimated the mass accretion rate of the Sgr A* to be $(1.7-7.0)\times10^{-8}\,M_{\odot}\,\mathrm{yr^{-1}}$ and $3.0\times10^{-8}\,M_{\odot}\,\mathrm{yr^{-1}}$ using an MHD and a GRMHD calculation, respectively. Using the model of Ref.~\cite{ozel2000hybrid}, Ref.~\cite{agol2000sagittarius} found that the upper limit of the accretion rate of the ADAF is about $3.0\times10^{-6}\,M_{\odot}\,\mathrm{yr^{-1}}$. Faraday rotation measurements provide upper limits of accretion rate of ADAF or a convection-dominated accretion flow of the Sgr A* to be $2\times10^{-7}\,M_{\odot}\,\mathrm{yr^{-1}}$ in ordered, radial, and equipartition magnetic field condition and lower limits to be $2\times10^{-9}\,M_{\odot}\,\mathrm{yr^{-1}}$ with subequipartition and disordered magnetic field ~\cite{marrone2006unambiguous}. Bower $\mathit{et}$ $\mathit{al.}$ also estimated $\dot{M}\sim10^{-8}\,M_{\odot}\,\mathrm{yr^{-1}}$ \cite{bower2018alma}. So in our model, we use the mass accretion rate in $=1.0\times10^{-8}\,M_{\odot}\,\mathrm{yr^{-1}}$ from Bower $\mathit{et}$ $\mathit{al.}$, which is $\dot{m}=1.13\times10^{-7}$.
We choose the ADAF parameters $\alpha=0.3$, $\beta=0.5$ (exact equipartition between the pressure of gas and magnetic field) and $f=0.9994$. The parameters are referred from Ref.~\cite{narayan1998advection} that applied the ADAF model to Sgr A*.

We calculate the creation rate of the neutron from nuclear reactions and the decay of the neutrons within the ADAF with chosen parameters and determine the number density of the produced neutrons.

\subsection{Neutron creation rate}
If neutrons are produced by a certain nuclear reaction, then the number of reactions per unit time per unit volume, i.e., the reaction rate $R_{\mathrm{reaction}}$, becomes the creation rate of the neutrons. Even if several neutrons are produced by a certain reaction, one just multiplies with the number of produced neutrons. We use the nonrelativistic reaction rate to produce neutrons
\begin{equation}
\label{reaction1}
R_{\mathrm{reaction},AB} = n_{A}n_{B}{\left\langle \sigma v \right\rangle}_{AB},
\end{equation}
where
\begin{eqnarray}
\label{reaction2}
{\left\langle \sigma v \right\rangle}_{AB} &=& \sqrt{\frac{8}{\pi m(k_{\mathrm{B}}T)^{3}}}\nonumber\\
& & \times\int_{0}^{\infty} {\sigma_{AB}(E)}E\,\mathrm{exp}\left(-\frac{E}{k_{\mathrm{B}}T}\right)\, dE,
\end{eqnarray}
and $A$, $B$, $\ldots$ are particle species. If $A=B$, then one should divide Eq. \eqref{reaction1} by 2. Thus, the creation rate of the neutron is
\begin{equation}
    \dot{n}_{C}=\sum_{k}^N (R_{\mathrm{reaction},AB})_{k},
\end{equation}
where $k$ corresponds to each nuclear reaction that creates neutrons. For precise calculation, one should use the relativistic reaction rate \cite{weaver1976reaction}. However, in our ADAF model, the maximum ion temperature is about a few $10^{12}\,\mathrm{K}$ ($k_{B}T/m_{p}c^{2}\lesssim0.09$). There is little difference between relativistic and nonrelativistic reaction rates at this temperature. Therefore, we use the nonrelativistic Maxwell-Boltzmann distribution as in Eq. \eqref{reaction2}.

In the case of the accretion disk around the solar mass black hole or neutron star, we should consider the secondary reactions of particles that are produced by the first nuclear reactions. But in the Sgr A* case, the produced particle number density is much lower than the ion number density of the disk. So we only consider the first nuclear reactions.

\begin{figure}
\includegraphics[width=8.66cm]{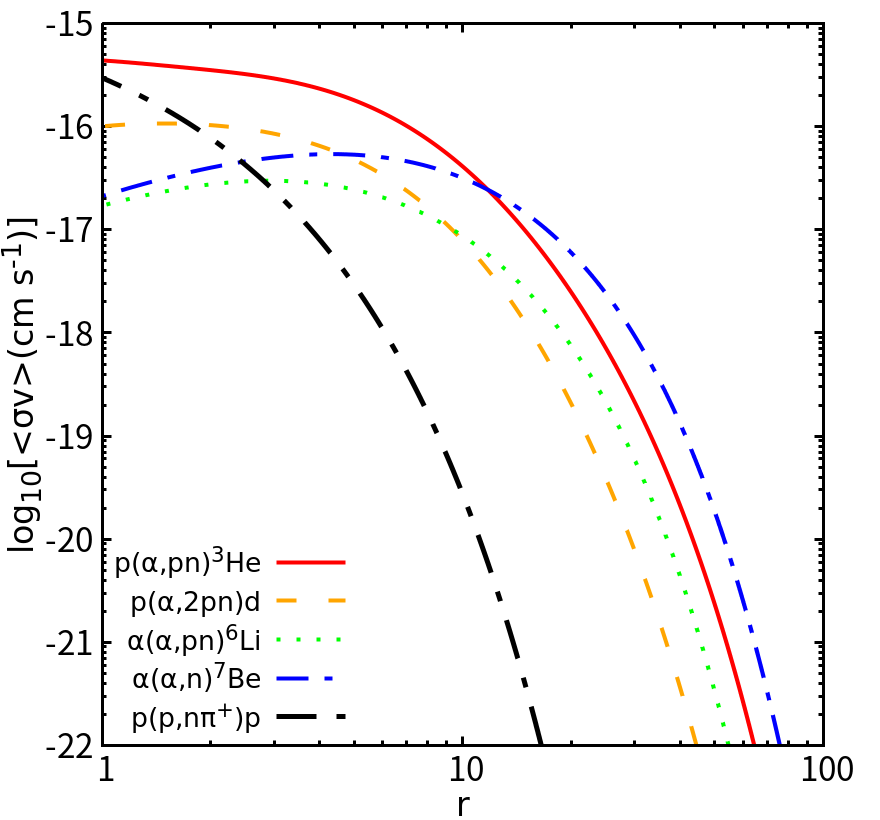}
\caption{The reaction rate ${\left\langle \sigma v \right\rangle}$ of $p(\alpha,pn)\mathrm{{}^{3}He}$ (red solid), $p(\alpha,2pn)d$ (orange dashed), $\alpha(\alpha,pn)\mathrm{{}^{6}Li}$ (green dotted), $\alpha(\alpha,n)\mathrm{{}^{7}Be}$ (blue dash dotted), and $p(p,n\pi^{+})p$ (black long dash dotted) are shown, respectively. The radial range is $1<r<100$. We use the ADAF parameters with $\alpha=0.3$, $\beta=0.5$, and $f=0.9994$.
}\label{fig1}
\end{figure}

In our model, we consider that protons and $\alpha$ particles ($^{4}\mathrm{He}$) dominate the accretion flow and consider reactions $p(\alpha,pn)\mathrm{{}^{3}He}$, $p(\alpha,2pn)d$, $\alpha(\alpha,pn)\mathrm{{}^{6}Li}$, $\alpha(\alpha,n)\mathrm{{}^{7}Be}$, and $p(p,n\pi^{+})p$. Also, we choose the mass fractions $X_{p}=0.75$ and $X_{\alpha}=0.25$ ~\cite{narayan1995advection}. Other reactions are excluded because their reaction cross section is much smaller than the above reactions. We use the cross section of $p-\alpha$ reactions from Refs.~\cite{meyer1972deuterons,jung1973proton}, $\alpha-\alpha$ reactions from Refs.~\cite{king1977li,yiou1977cross,woo1985cross,mercer1997suggested,mercer2001production,kawabata2017time} and $p(p,n\pi^{+})p$ reaction from Ref.~\cite{kelner2006energy}. 
Figure \ref{fig1} shows ${\left\langle \sigma v \right\rangle}$ of these reactions in our model as a function of $r$.

\subsection{Neutron inflow}
In general, ions and electrons cannot escape from the accretion disk and are accreted to the central object because of the electromagnetic interaction. But neutrons are free from the electromagnetic force and easily escape from the disk. Thus, it is important to estimate how many neutrons do not escape to infinity and flow into the acceleration zone close to the black hole. 

Neutrons can be confined by the gravitational binding and the scattering with ions \cite{guessoum1990neutron}. Neutrons are scattered with ions with the mean free path
\begin{equation}
    \lambda_{n}=\frac{1}{n_{i}\sigma},
\end{equation}
where $n_{i}$ is an ion number density and $\sigma$ is a cross section with ions.
In the ADAF disk of the Sgr A*, the mean free path of the neutrons produced within $r\sim10$ is longer than $r\sim10^{4}$. Therefore, neutrons are not affected by the plasma in the disk and can be treated as collisionless.

Aharonian and Sunyaev \cite{aharonian1984gamma} presented an escape fraction of the neutrons produced in an accretion disk. They used the condition that if the sum of the thermal energy of neutrons and the energy from their flow velocity exceeds the gravitational binding energy, then neutrons can freely escape to infinity:
\begin{equation}
\label{neutronev1}
    \frac{1}{2}m_{\mathrm{n}}(v_{\mathrm{flow}}+v_{\mathrm{th}})^{2}\geq \frac{GMm_{\mathrm{n}}}{R}.
\end{equation}

From this condition, the escape fraction of neutrons is
\begin{eqnarray}
\label{fraction}
    f_{esc}=\left(\frac{m}{2\pi k_{B}T_{i}}\right)^{3/2}\int_{0}^{2\pi}\,d\phi\int_{0}^{\pi}\mathrm{sin}\theta\, d\theta\nonumber\\
    \times\int_{v_{min}}^{\infty}v_{th}^{2}e^{-mv_{th}^{2}/2k_{B}T}\,dv_{th}.
\end{eqnarray}
using the Maxwell-Boltzmann distribution.

But, our purpose is to estimate the inflow rate of neutrons into the magnetic field near the black hole. So we integrate fractions that flow into the magnetic field using geodesic integration. To be exact, we need to know the real neutron velocity distribution which may be different from the Maxwell-Boltzmann distribution. However, there is no experimental data on the neutron energy spectrum of nuclear reactions that are used in our calculation, and there are more than three products in the reactions. Therefore, we assume the neutron energy spectrum to be the Maxwell-Boltzmann distribution of ion temperature $T_i$ with the following assumptions. 

When a neutron and a $X$ particle are produced by a nuclear reaction, momentum conservation at the center of mass is
\begin{equation}
\label{moconv}
    m_n\vec{u}_n+m_X\vec{u}_X=0,
\end{equation}
where $\vec{u}$ is the velocity with respect to the center of mass. Energy conservation requires
\begin{equation}
\label{econv}
    Q+K=\frac{1}{2}m_n{u_n}^2+\frac{1}{2}m_X{u_X}^2,
\end{equation}
where $Q$ is the mass difference before and after the reaction and $K$ is the kinetic energy of reactants at the center of mass. If there are multiple $N$ particles, then one needs the relative velocity of each of them. However, if $N$ is greater than two, this is unknown. So, we apply following assumptions to Eqs. \eqref{moconv} and \eqref{econv}:
\begin{eqnarray}
    m_n\vec{u}_n+\sum^{N}_{i=1}(m_i)\vec{u}_\mathrm{c.m.}=0,\qquad\ \nonumber\\
    Q+K\approx\frac{1}{2}m_n{u_n}^2+\frac{1}{2}\sum^{N}_{i=1}(m_i){u_\mathrm{c.m.}}^2,
\end{eqnarray}
where $\vec{u}_\mathrm{c.m.}$ is the center of mass velocity of $N$ particles. Under this condition, the  mean energy of neutrons \cite{brysk1973fusion} is
\begin{equation}
    \left\langle E_n \right\rangle=\frac{1}{2}m_n\left\langle V^2 \right\rangle+\frac{\sum^{N}_{i=1}(m_i)}{m_n+\sum^{N}_{i=1}(m_i)}\left(Q+\left\langle K \right\rangle\right),
\end{equation}
where $V$ is the total center of mass velocity and
\begin{eqnarray}
    \left\langle V^2 \right\rangle=\frac{3k_BT_i}{m_n+\sum^{N}_{i=1}(m_i)},\quad\quad\ \nonumber\\
    \left\langle K \right\rangle=\frac{\int_{0}^{\infty} K^2dK\sigma(K)\mathrm{exp}(-K/k_BT_i)}{\int_{0}^{\infty} KdK\sigma(K)\mathrm{exp}(-K/k_BT_i)}.
\end{eqnarray}
For the dominant nuclear reaction $p(\alpha,pn)\mathrm{{}^{3}He}$ within $r\lesssim10$, the mean neutron energy $\left\langle E_n \right\rangle$ and the mean energy of Maxwell-Boltzmann distribution with $T_i$ are almost identical. Therefore, we assume $T_i$ to be the neutron temperature. 

Neutrons decay with a lifetime. So, the fraction of neutrons reaching a certain radius $r$ decreases as the travel time increases. We calculate the neutron travel time $\tau$ from the geodesic equation.
\begin{equation}
\frac{du^{\beta}}{d\tau}=-\Gamma^{\beta}_{\mu\nu}u^{\mu}u^{\nu},
\end{equation}
where the Christoffel symbol is
\begin{equation}
    \Gamma^{\beta}_{\mu\nu}=\frac{1}{2}g^{\beta\alpha}(\partial_{\nu}g_{\alpha\mu}+\partial_{\mu}g_{\alpha\nu}-\partial_{\alpha}g_{\mu\nu}).
\end{equation}
Since self-similar ADAF solutions, used in our model, are based on Newtonian gravity and Sgr A* has a relatively small spin parameter, we calculate the motions of neutrons with Schwarzschild metric as a leading order. The Schwarzschild metric for a black hole of mass $M$ is
\begin{eqnarray}
    ds^{2}&=&-c^{2}d\tau^{2}\nonumber\\
    &=&-\left(1-\frac{R_{s}}{R}\right)c^{2}dt^{2}+\left(1-\frac{R_{s}}{R}\right)^{-1}dR^{2}\nonumber\\
    & &+R^2\left(d\theta^2+\mathrm{sin}^{2}\theta d\phi^{2}\right).
\end{eqnarray}

Self-similar ADAF solutions provide only radial velocity and angular frequency profiles of the accretion flow without the vertical flow structure information \cite{narayan1995advection}. Therefore, we vertically average the number of neutrons. From the assumption, the total neutron number at initial radius $r_i$ is
\begin{eqnarray}
    N_{n,i}(r_i)&=&\int n_{n,i}\, dV\sim n_{n,i}(r_i)\Delta V\nonumber\\
    &=&4\pi H{R_s}^2n_{n,i}(r_i)r_i\Delta r,
\end{eqnarray}
where $2\pi$ is from the axial symmetry of the self-similar ADAF model.

We calculate neutron inflow using the geodesic of neutrons. The neutron density at initial position $\vec{r}_i=(r_i,\pi/2)$ with an initial velocity on the local rest frame of the accretion flow $\vec{v}_i=({v_i}^{(r)},{v_i}^{(\theta)},{v_i}^{(\phi)})=({v_i}\mathrm{cos}\theta',{v_i}\mathrm{sin}\theta'\mathrm{cos}\phi',{v_i}\mathrm{sin}\theta'\mathrm{sin}\phi')$ is
\begin{eqnarray}
    n_{n,i}(\vec{v}_i,r_i)&\sim&\dot{n}_c(r_i)\frac{R_s\Delta r}{\left|{v_r}(r_i)\right|}\left(\frac{m_n}{2\pi k_B T_i}\right)^{3/2}\nonumber\\
    & &\times {v_i}^2e^{-m{v_i}^2/2k_BT_i}\mathrm{sin}\theta'\Delta\theta'\Delta\phi'\Delta v_i,
\end{eqnarray}
where $\theta'$ and $\phi'$ are from the coordinate of the local rest frame, and ${v_r}(r_i)$ is the proper radial velocity of the accretion flow at an initial position. We choose the ADAF radial velocity of the Eq. \eqref{ADAF1} to be the proper radial velocity of an accretion flow at an initial position.

The neutron number of the $k\mathrm{th}$ step at $\vec{r}_k=(r_k,\theta_k)$ is
\begin{equation}
    N_{n,k}(\vec{v}_i,\vec{r}_{k},r_i)= N_{n,i}(\vec{v}_i,r_i)\exp\left[-\frac{\tau (\vec{v}_i,\vec{r}_{k},r_i)}{\tau_n}\right],
\end{equation}
where $\tau$ the neutron travel time from $r_i$ to $\vec{r}_{k+1}$, the neutron life time $\tau_n$ is about 879.4 seconds \cite{particle2020review}.

Therefore, the total neutron number at the final position $\vec{r}_f$ is
\begin{equation}
    \label{nnrf}
    \Delta N_n(\vec{r}_f)=\sum_{r_i}\sum_{\vec{v}_i}N_{n,f}(\vec{v}_i,\vec{r}_f,r_i)
\end{equation}
Figure \ref{fig1} shows that the neutron production reactions become efficient only at $r\lesssim10$. Thus, for our calculation, we set $1<r<10$ with $\Delta r=5\times10^{-3}$. We also set that $\Delta\theta'=\pi/30$, $\Delta\phi'=2\pi/30$, and $\Delta v_i=3\sqrt{k_BT_i/m_n}/100$ with $0<\theta'<\pi$, $0\leq\phi'<2\pi$, and $0<v<3\sqrt{k_BT_i/m_n}$, respectively.

We compare the escape fraction of our model with that of Aharonian and Sunyaev \cite{aharonian1984gamma}. For $T_i\approx1.3\times10^{11}\,\mathrm{K}$, our model shows that about $18\%$ of neutrons at $R\sim(7-9)R_s$ escape to infinity when neutron decay is not considered, which is almost same as the estimate of Aharonian and Sunyaev. However, the difference between the two models will be larger much closer to the black hole by general relativistic effect.

\section{Proton acceleration through the MPP}

In this chapter, we present the accelerated proton energy spectrum from the MPP in the Kerr space-time. The Kerr metric for a black hole of mass $M$ and an angular momentum of $J$ is 
\begin{eqnarray}
\label{Kerr}
    ds^{2}&=&-c^{2}d\tau^{2}\nonumber\\
    &=&-\left(1-\frac{R_{s}R}{\Sigma}\right)c^{2}dt^{2}+\frac{\Sigma}{\Delta}dR^{2}+\Sigma d\theta^{2}\nonumber\\
    & &+\left(R^{2}+a^{2}+\frac{R_sRa^{2}}{\Sigma}\mathrm{sin}^{2}\theta\right)\mathrm{sin}^{2}\theta d\phi^{2}\nonumber\\
    & &-\frac{2R_{s}R a\mathrm{sin}^{2}\theta}{\Sigma}cdtd\phi
\end{eqnarray}
in the Boyer–Lindquist coordinates, where $\Sigma$, $\Delta$, and $a$ are
\begin{eqnarray}
    \Sigma=R^{2}+a^{2}\mathrm{cos}^{2}\theta,\qquad\quad \nonumber\\
    \Delta=R^{2}-R_{s}R+a^{2},\quad a=\frac{J}{Mc},
\end{eqnarray}
\label{Kerr2}
respectively.

\begin{figure}
\includegraphics[width=8.65cm]{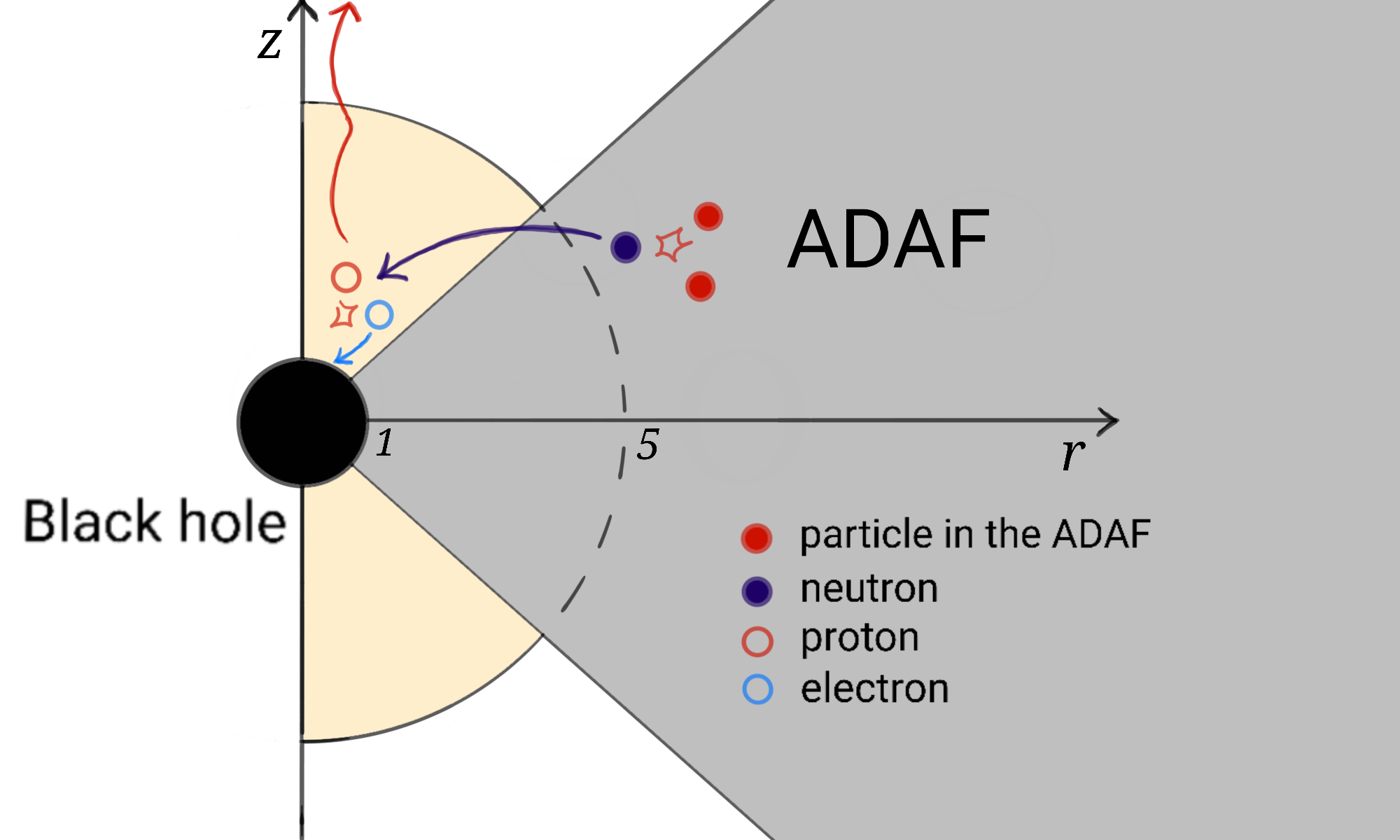}
\caption{The schematic view of our model at the vicinity of a black hole is described by cylindrical coordinates. Neutrons are produced by nuclear reactions in the accretion disk (the region of gray color) and flow into the acceleration zone (the region of beige color).}
\label{Scheme}
\end{figure}

In the MPP model of Tursunov $\mathit{et}$ $\mathit{al.}$ (2020), protons that are produced by neutron decay are accelerated through electric potential difference \cite{tursunov2020supermassive}. If protons are produced by neutron decay in an accretion disk, then it is challenging to accelerate protons because plasma in the accretion disk can screen the electric field induced by the rotation of a black hole and an external uniform magnetic field. In such a case,  if the charged particle density of plasma is lower than the Goldreich-Julian density, acceleration of charged particles can operate \cite{rueda2022gravitomagnetic,goldreich1969pulsar,akiyama2022first}. However, the charged particle density of the accretion disk of the Sgr A* from Eq. \eqref{ADAF3} is $n_{Sgr A*}\sim 10^{6}-10^{5}\,\mathrm{cm}^{-3}$ at $r\lesssim10$ and the Goldreich-Julian density of the Sgr A* is $n_{GJ}\sim a_*B_0c^2/\left(4\pi eGM\right)\sim10^{-2}a_*\left[B_0/\left(30 \,\mathrm{G}\right)\right]\,\mathrm{cm}^{-3}$ \cite{akiyama2022first}, where $a_{*}=2a/R_s$ is a dimensionless spin parameter. So the proton acceleration after neutron decay by the induced electric field is difficult in the accretion flow of Sgr A* and does not operate inside the accretion flow. We assume that there is no plasma in the polar region of the accretion flow close to Sgr A* from estimate to accretion flow more likely due to the outflow \cite{narayan1995advection2,park1999thermal,park2007compton}. The electron-positron pair density is estimated to be $n_{\pm}\approx10^{-8}\,\mathrm{cm^{-3}}$ which is much lower than the Goldreich-Julian density in the GRMHD simulations \cite{moscibrodzka2011pair,wong2021pair} and will not screen the electric field. Therefore, we set the conditions of the acceleration zone to operate MPP are $0<\theta<\theta_{cut,1}=\pi/2-\mathrm{tan}^{-1}\left(H/R\right)$, $\theta_{cut,2}=\pi/2+\mathrm{tan}^{-1}\left(H/R\right)<\theta<\pi$ and $1<r<5$. We assume that protons from neutrons decay are accelerated and escape without any hindrance of the plasma in the acceleration zone. We express the schematic view of our model in Fig. \ref{Scheme}.

Tursunov $\mathit{et}$ $\mathit{al.}$ (2020) calculated the maximum efficiency $\eta_{\mathrm{ultra}}$ of a proton that is produced in the vicinity of the black hole by a neutron decay with Wald solution:
\begin{equation}
    \eta_{\mathrm{ultra}}=\eta_{\mathrm{PP}}+\frac{q}{m}A_{t},\quad \eta_{\mathrm{PP}}\simeq0.21,
\end{equation}
where $q$ is a charge of the proton, $m$ is a mass of the neutron, and $A_{t}$ is a covariant electric potential. At this efficiency, the energy of the accelerated proton is
\begin{equation}
    E_{\mathrm{p}}=1.7 \times 10^{20}\,\mathrm{eV}\left(\frac{B}{10^{4}\,\mathrm{G}}\right)\left(\frac{M}{10^{9}M_{\odot}}\right)\left(\frac{a_{*}}{0.8}\right).
\end{equation}
In the Sgr A* case, this energy is $E_{p}\simeq5\times10^{15}\,\mathrm{eV}$.

\subsection{Wald solution}
Wald solution is the vacuum Maxwell equation solution when a Kerr black hole is immersed in a uniform magnetic field. Although Wald solution is a vacuum solution, for the condition that the density of plasma near black hole $n\ll n_{GJ}$ and the large-scale magnetic field shares the axis symmetry and stationary of space-time geometry, Wald solution has been employed to demonstrate the motion and the acceleration of charged particles in the vicinity of a black hole to leading order in many studies \cite{wald1974black, stuchlik2016acceleration,tursunov2016circular,kolovs2017possible,ruffini2019gev,rueda2020blackholic,tursunov2020supermassive,moradi2021newborn,rueda2022gravitomagnetic,rastegarnia2022structure}. The electromagnetic vector field presented by the Wald solution ($G=c=1$) is
\begin{equation}
\label{Killing2}
    A_{\mu}=\frac{1}{2}B_{0}\left(\psi_{\mu}+\frac{2J}{M}\eta_{\mu}\right)-\frac{Q}{2M}\eta_{\mu},
\end{equation}
where $B_{0}$, $J$, and $Q$ are a uniform magnetic field, an angular momentum of the black hole, and a charge of the black hole, and $\psi_{\mu}$ and $\eta_{\mu}$ Killing vectors that respond to $\partial/\partial\phi$ and $\partial/\partial t$, respectively. Wald solution, expressed by metric tensor, is
\begin{equation}
    A_{\mu}=\frac{1}{2}B_{0}\left(g_{\mu\phi}+2ag_{\mu t}\right)-\frac{Q}{2M}g_{\mu t}.
\end{equation}
When the charge $Q=0$, a contravariant electric vector field $A^{t}$ is nonzero. Thus, the vector field raises the selective charge accretion in a very short time in astrophysical black holes \cite{kolovs2017possible,tursunov2020supermassive}. When the black hole charge $Q=2aMB_{0}$ which is a Wald charge, the black hole finishes the selective accretion. Thus, we can write the vector potential as
\begin{equation}
\label{Killing4}
    A_{\mu}=\frac{1}{2}B_{0}g_{\mu\phi}.
\end{equation}
From the short timescale of selective charge accretion of an astrophysical black hole, a scenario that a black hole is induced with a Wald charge is applied to describe the motion of charged particles. For gamma-ray burst cases where a black hole has just been born, the acceleration of charged particles is estimated from Wald solution with charge $Q=0$, and electrons escape to infinity and protons fall into a black hole at particular direction \cite{ruffini2019gev,moradi2021newborn,rueda2022gravitomagnetic,rastegarnia2022structure}.
When a black hole has a Wald charge and decaying neutrons nearby, the proton escape scenario is preferred because it is highly likely that the black hole will be charged with a positive Wald charge \cite{wald1974black,zajavcek2018charge,zajavcek2019electric}. Thus, we adopt the scenario that electrons fall into the black hole and protons escape to infinity after neutrons decay.
When protons escape from the vicinity of the black hole to infinity, they have two components of energy, which are the vertical escape energy $E_{z}$ and oscillatory (Larmor) energy in the equatorial plane $E_{L}$.
\begin{equation}
    E^{2}_{\infty}=E^{2}_{z}+E^{2}_{L}.
\end{equation}
If a black hole is induced with a Wald charge, then the vertical energy of charged particles is maximized, and the charged particles escape to infinity in a vertical direction \cite{tursunov2020supermassive,stuchlik2016acceleration}.

\subsection{The production rate of the accelerated protons}
First, we assume that the Galactic center black hole, Sgr A*, is induced with the Wald charge and protons just escape to infinity without falling into the black hole with maximized vertical escape energy. Since the Kerr metric has two Killing vectors $\psi_{\mu}$ and $\eta_{\mu}$, the proton energy and angular momentum are conserved. The conserved total energy of the proton is
\begin{equation}
    E=-\left(mu_{t}\right)_{\mathrm{decay}}-qA_{t,\mathrm{decay}}=-\left(mu_{t}\right)_{\infty}-qA_{t,\infty},
\end{equation}
where $-\left(mu_{t}\right)_{\mathrm{decay}}$ and $-\left(mu_{t}\right)_{\infty}$ are proton energies at a decay point and infinity, respectively, and $-qA_{t}$ is electric potential energies at corresponding locations. Since protons escape vertically, and we can ignore electric field $A_{t,\infty}$ ($A_{t,\infty}\ll A_{t,\mathrm{decay}}$), the proton energy at infinity is
\begin{equation}
    E_{\mathrm{p}}=-\left(mu_{t}\right)_{\infty}\simeq-\left(mu_{t}\right)_{\mathrm{decay}}-qA_{t,\mathrm{decay}}.
\end{equation}
Since the proton is accelerated in a vertical direction regardless of the proton's motion at the decay point and $\vert-\left(mu_{t}\right)_{\mathrm{decay}}\vert \ll\left\vert-qA_{t,\mathrm{decay}}\right\vert$, one can write the proton energy at infinity as
\begin{equation}
    E_{\mathrm{p}}\simeq-qA_{t,\mathrm{decay}}.
\end{equation}
Thus, when we adopt the Eq. \eqref{Killing4} for $A_{t,\mathrm{decay}}$ ($c\neq G\neq1$),
\begin{equation}
\label{protonem1}
    E_{p}=-\frac{1}{2}qB_{0}cg_{t\phi}=\frac{qB_{0}c}{2}\frac{R_{s}r a\,\mathrm{sin}^2\theta}{r^{2}+a^{2}\mathrm{cos}^{2}\theta}.
\end{equation}
The proton production rate $Q_{MPP}\left(E_p\right)$ is now
\begin{equation}
    Q_{MPP}=\frac{d\dot{N}_p}{dE_p}\sim\frac{\Delta\dot{N}_p}{\Delta E_p}=\frac{\Delta N_n}{\tau_n\Delta E_p},
\end{equation}
where $\Delta E_p$ is the total differential of the accelerated proton,
\begin{equation}
    \Delta E_p=\frac{\partial E_p}{\partial r}\Delta r+\frac{\partial E_p}{\partial \theta}\Delta\theta.
\end{equation}
However, $\Delta N_n$ and $\Delta E_p$ depend on $r$ and $\theta$. Thus, we calculate the total proton production rate after dividing the acceleration zone with radius $r$ and calculating each production rate. For a given $r$,
\begin{eqnarray}
    E_p&=&E_{p,r}\left(\theta\right),\quad\Delta N_n=\Delta N_{n,r}\left(\theta\right),\nonumber\\ 
    & &\Delta E_{p,r}\left(\theta\right)=\left.\frac{\partial E_p}{\partial \theta}\right|_r\left(\theta\right)\Delta \theta.
\end{eqnarray}
Proton production rate at $r\sim r+\Delta r$ is
\begin{equation}
    Q_{MPP,r}\left(E_{p,r}\right)\sim\frac{\Delta N_{n,r}\left(\theta\right)}{\tau_n\Delta E_{p,r}\left(\theta\right)}.
\end{equation}
Therefore, the total production rate is
\begin{equation}
\label{Qmpp}
    Q_{MPP}\left(E_p\right)=\sum_{r}Q_{MPP,r}\left(E_p\right).
\end{equation}
We set $\Delta r=0.01$ and $\Delta \theta=\pi/100$. 

We select the dimensionless spin parameter and uniform magnetic field strength in the vicinity of the black hole as $a_{*}=0.5$ and $B_{0}=30,\,100\,\mathrm{G}$ \cite{tursunov2020supermassive}. The spin parameter value corresponds to orbiting spot models with variable emission \cite{eckart2017milky}. The magnetic field strength coincides with the emission model of Sgr A*, which requires the ordered magnetic field with $B_{0}=30-100\,\mathrm{G}$ and most accretion models in the equipartition state have the magnetic field with a few hundred Gauss \cite{eatough2013strong}. For example, self-similar solutions that are used in our model estimate the magnetic field of a few hundred Gauss \cite{narayan1995advection}, and RIAF based on Keplerian shell model with radio emission proposed the magnetic field strength to about 100 Gauss at $r\sim1$ \cite{cho2022intrinsic}.

\begin{figure}
\includegraphics[width=8.66cm]{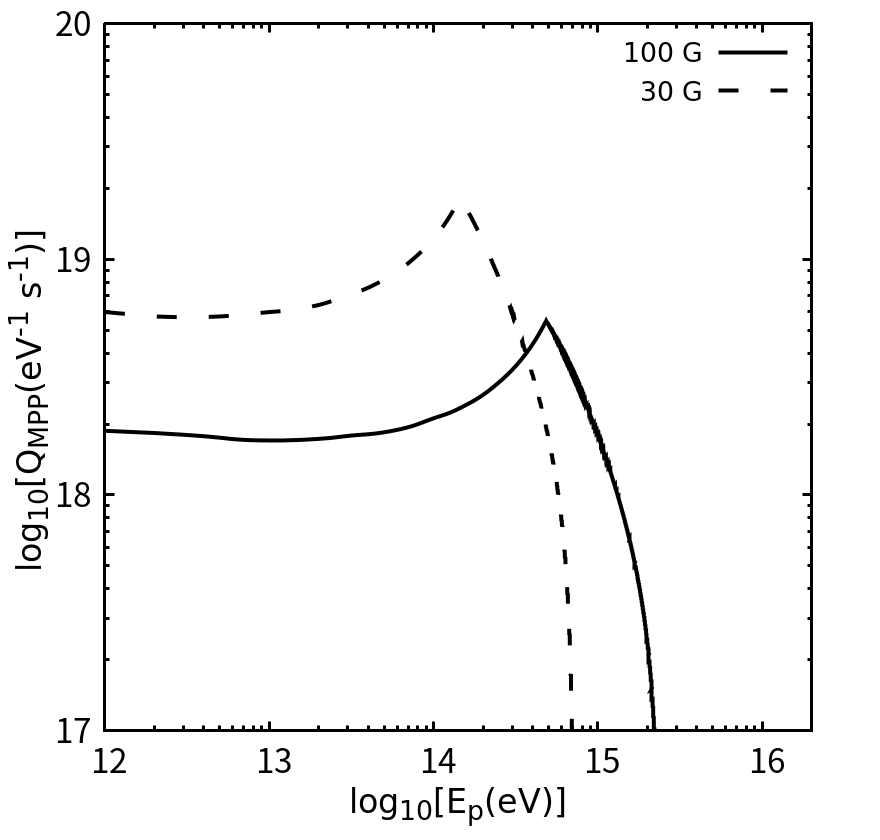}
\caption{The production rates of protons that accelerated by MPP from Eq. \eqref{Qmpp}. The solid and dashed lines show the proton production rates for $100$ and $30\,\mathrm{G}$.}
\label{production}
\end{figure}
From Eq. \eqref{Qmpp}, we get the production rate of protons accelerated by MPP as shown in Fig. \ref{production}. At the neighborhood of the polar axis ($\lesssim10^{14}\,\mathrm{eV}$ for $B_0=100\,\mathrm{G}$ and $\lesssim10^{13}\,\mathrm{eV}$ for $B_0=30\,\mathrm{G}$), $\Delta N_n$ and $\left.\partial E_p/\partial \theta\right|_r$ increase similarly as $\theta$ approaches $\theta_{cut}$. At the neighborhood of the $\theta_{cut}$ ($\gtrsim10^{14}\,\mathrm{eV}$ for $B_0=100\,\mathrm{G}$ and $\gtrsim10^{13}\,\mathrm{eV}$ for $B_0=30\,\mathrm{G}$), $\Delta N_n$ increase faster than $\left.\partial E_p/\partial \theta\right|_r$ as $\theta$ approaches $\theta_{cut}$. From those feature of the neutron number and the partial derivative of the energy of protons, the production rate of protons increases from about $10^{14}\,\mathrm{eV}$ for $B_0=100\,\mathrm{G}$ (about $10^{13}\,\mathrm{eV}$ for $B_0=30\,\mathrm{G}$) to the peak energy $E_{peak}\simeq 4.8\times10^{14}\,\mathrm{eV}$ ($E_{peak}\simeq 1.5\times10^{14}\,\mathrm{eV}$ for $B_0=30\,\mathrm{G}$). The production rate of protons at $E_{peak}$ is $5.5\times10^{18}\,\mathrm{eV^{-1}\,s^{-1}}$ for $B_0=100\,\mathrm{G}$ and $1.8\times10^{19}\,\mathrm{eV^{-1}\,s^{-1}}$ for $B_0=30\,\mathrm{G}$. $E_{peak}$ is the energy of protons accelerated at $r=5$ and $\theta=\theta_{cut}$. As the radius where the proton is accelerated decreases, protons are accelerated to an energy higher than $E_{peak}$ on $r<5$. Therefore production rate of protons steeply decreases from $E_{peak}$.

\section{Emission from accelerated protons}

\subsection{Synchrotron}
The main radiation processes of high-energy charged particles are synchrotron, inverse Compton scattering and bremsstrahlung.

Charged particles have chaotic motion when gravity and electromagnetic force are strong. One can check which force more strongly affects charged particles through the parameter
\begin{equation}
    \mathcal{B}=\frac{qGMB}{2mc^{3}}.
\end{equation}
When $\mathcal{B}<1$, charged particles fall freely into the black hole, while $\mathcal{B}\sim1$, the motion of charged particles is chaotic. For $\mathcal{B}\gg1$, the electromagnetic effect on charged particles is greater than gravity, and the charged particles escape to the direction parallel to the rotational axis of the black hole and magnetic field. For larger $\mathcal{B}$, charged particles are accelerated faster and the pitch angle goes to zero \cite{tursunov2020supermassive,stuchlik2016acceleration,tursunov2016circular,kolovs2017possible,stuchlik2021penrose}. Since the pitch angle is close to zero, synchrotron radiation of accelerated protons by MPP in the vicinity of Sgr A* will be minimized. We can simply compare the synchrotron radiation and luminosity of Sgr A*. Without gravitational redshift, the power spectrum of synchrotron radiation of a charged particle is \cite{rybicki1991radiative}
\begin{eqnarray}
\label{pnu}
    P_{\nu}&=&\frac{\sqrt{3}q^{3}B}{mc^{2}}F(\nu/\nu_{c})\mathrm{sin}\psi,\nonumber\\F(x)&=&x\int_{x}^{\infty} K_{5/3}(y)\, dy,
\end{eqnarray}
where $\nu_{c}$ is the characteristic frequency of synchrotron radiation,
\begin{equation}
    \nu_{c}=\frac{3\gamma^{2}qB}{4\pi mc}\mathrm{sin}\psi.
\end{equation}
For $B=100\,\mathrm{G}$ and proton with $E=0.48\,\mathrm{PeV}$, the characteristic frequency is $6.24\times10^{18}\mathrm{sin}\psi\,\mathrm{Hz}$, and the characteristic energy of the photon is $258\mathrm{sin}\psi\,\mathrm{eV}$. If one assumes that protons escape the vicinity of Sgr A* with velocity $v\sim c$ and parallel direction to the uniform magnetic field and the rotational axis of Sgr A*, from Eqs. \eqref{Qmpp} and \eqref{pnu}, then at the peak frequency $\sim2\times10^{15}\,\mathrm{Hz}$, $\left(\nu L_{\nu}\right)_{peak}\sim10^{26}\,\mathrm{erg}\,\mathrm{s^{-1}}$ on the scale of the Schwarzschild radius of Sgr A*. We estimated $(\nu L_{\nu})_{peak}$ only on the scale of the Schwarzschild radius, taking into account the decrease in the magnetic field with distance from Sgr A*. However, the best-fitting power law from the mean NIR to x-ray spectra of VB3 which was observed on 30 August 2014 and is one of the very bright flares of Sgr A* shows that $\nu L_{\nu}\sim10^{35}\,\mathrm{erg\,s^{-1}}$ at $\nu\sim10^{15}\,\mathrm{Hz}$ \cite{ponti2017powerful}. Also RIAF models show $\nu L_{\nu}\sim10^{33}\,\mathrm{erg\,s^{-1}}$ at $\nu\sim10^{15}\,\mathrm{Hz}$ \cite{narayan1998advection,yuan2004nature}. The small pitch angle will further decrease the synchrotron emission, and synchrotron radiation from accelerated proton does not affect the spectrum of Sgr A*.

\subsection{Inverse Compton scattering and bremsstrahlung}
The effect of inverse Compton scattering from the Compton $y$ factor. For monoenergetic proton, the Compton $y$ factor is 
\begin{equation}
    y=\frac{4}{3}\gamma^{2}\beta^{2}\times \mathrm{max}(\tau,\tau^{2}),\quad\tau=n_p\sigma_{T}\left(\frac{m_e}{m_p}\right)^{2}L,
\end{equation}
where $\tau$ is optical depth and $L$ is the length scale. On the scale of the Schwarzschild radius with $B_0=100\,\mathrm{G}$, the number density of accelerated protons
\begin{eqnarray}
\label{numdenp}
    n_p&\sim&\frac{1}{{R_s}^{3}}\cdot\frac{2R_s}{c}\cdot\int Q_{MPP}(E_p)\, dE_p\nonumber\\
    &\sim&\frac{2}{{R_s}^2c}\cdot Q_{MPP}\left(E_{peak}\right)E_{peak}\sim0.123\,\mathrm{cm^{-3}}
\end{eqnarray}
Assuming $L=R_s$, $\tau\sim2.9\times10^{-20}$. Therefore,
\begin{equation}
    y\sim\frac{4}{3}\cdot(4.8\times10^{5})^2\cdot2.9\times10^{-20}=8.9\times10^{-9}.
\end{equation}
For high-energy photons like x-rays, the Klein-Nishina scattering cross section $\sigma_{KN}$ should be used, and $\sigma_{KN}$ will make the Compton $y$ factor even smaller. With this small value of the Compton $y$ factor, inverse Compton scattering of accelerated protons will not affect the spectrum of Sgr A*.

The emission of photons by high-energy ions interact with electrons is called the inverse bremsstrahlung \cite{baring2000inverse}. The inverse bremsstrahlung against the electrons in the accretion flow is very low for cross section of inverse bremsstrahlung \cite{baring2000inverse} and the inverse bremsstrahlung neither will affect the spectrum of Sgr A*.

\subsection{Proton-proton interaction}
Many observations found that the central molecular zone (CMZ) in the GC produces very high energy (VHE) gamma rays (which is called HESS J1745-290 or VER J1745-290) \cite{collaboration2016acceleration,tsuchiya2004detection,kosack2004tev,aharonian2004very,albert2006observation,aharonian2009spectrum,abdalla2018characterising,acciari2020magic,adams2021veritas}. Inverse Compton scattering of high-energy electrons and p-p interaction of high-energy protons can produce the VHE gamma rays. HESS collaboration argues that the p-p interaction is more favorable than the inverse Compton scattering of high-energy electrons, in the case of the diffuse VHE emission of the CMZ based on the propagation scale of high-energy protons and electrons \cite{collaboration2016acceleration}. They also argue that one or more multi-TeV accelerators are in the CMZ based on VHE gamma ray observation data and gas distribution of the CMZ. 

If the accelerators inject high-energy particles at a continuous rate in the GC, then the radial distribution of the particles will be
\begin{equation}
    \frac{dn}{dE}=\frac{Q(E)}{4\pi D(E)r_{CMZ}},
\end{equation}
where $Q(E)$ is a injection rate of the particles and $D(E)$ is a diffusion coefficient. Therefore, using the energy spectrum of the protons which are accelerated from the MPP at the Eq. \eqref{Qmpp}, one can calculate the MPP's contribution to the gamma ray flux from the GC.

\begin{figure}
\includegraphics[width=8.66cm]{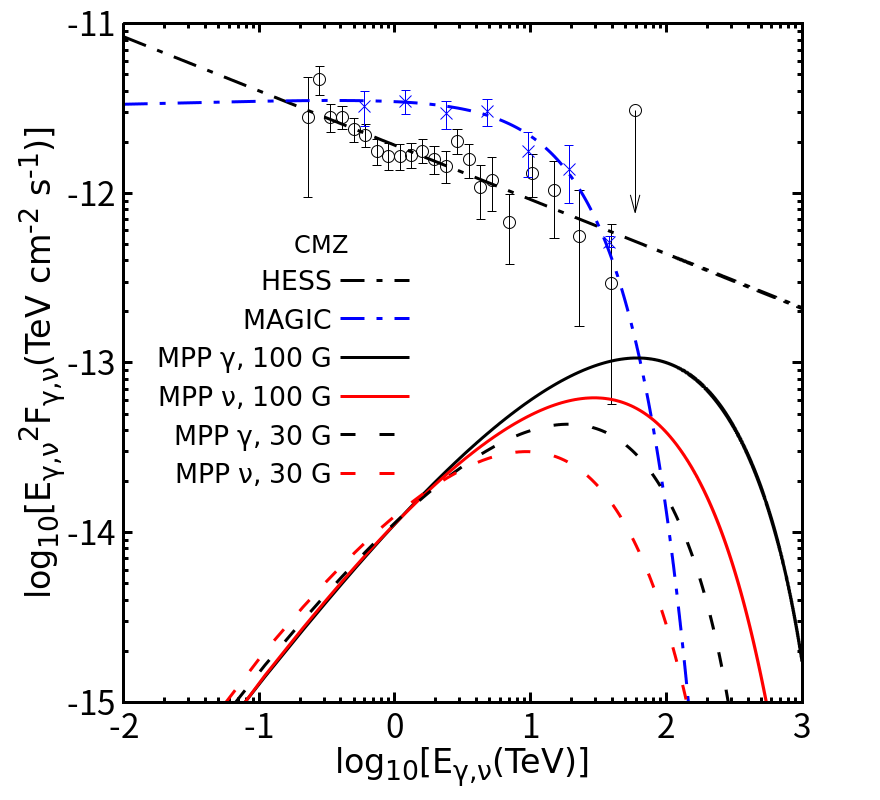}
\caption{Observation data of diffuse VHE gamma ray from CMZ (black open circles from HESS with the pac-man region \cite{collaboration2016acceleration} and blue crosses from MAGIC \cite{acciari2020magic}), and the fluxes of gamma rays and neutrinos from protons accelerated by the MPP in Sgr A*. Black lines and red lines represent fluxes of gamma ray and neutrinos respectively. Solid lines and dashed lines show gamma ray and neutrino fluxes in the cases of the magnetic field strength of the vicinity of the Sgr A* $B_0=100\,\mathrm{G}$ and $B_0=30\,\mathrm{G}$, respectively.}
\label{gammadif}
\end{figure}

The production rate of gamma rays and neutrinos from p-p interaction is
\begin{eqnarray}
    \frac{d\dot{n}_{\gamma, \nu}}{dE_{\gamma, \nu}}=\qquad\qquad\qquad\qquad\qquad\qquad\nonumber\\
    cn_\mathrm{H}\int_{E_{\gamma, \nu}}^{\infty} \sigma_{\mathrm{inel}}(E_p)\frac{dn_p}{dE_p}F_{\gamma, \nu}\left(\frac{E_{\gamma, \nu}}{E_p}, E_p\right)\frac{dE_p}{E_p},
\end{eqnarray}
where $n_\mathrm{H}$ is the density of the hydrogen gas, $\sigma_{\mathrm{inel}}(E_p)$ the inelastic cross section of the p-p interaction and $F_{\gamma, \nu}$ functions of the energy spectra of the gamma rays and neutrinos \cite{kelner2006energy}. Thus, the fluxes of the diffuse VHE gamma rays and neutrinos from the p-p interaction of protons that are accelerated by the MPP in the vicinity of the Sgr A* is
\begin{equation}
\label{fgn}
    F_{\gamma, \nu}\left(E_{\gamma, \nu}\right)=\frac{1}{4\pi \left(8\,\mathrm{kpc}\right)^{2}}\int  \frac{d\dot{n}_{\gamma, \nu}}{dE_{\gamma, \nu}}\left(E_{\gamma, \nu}\right)\, dV,
\end{equation}
where $8\,\mathrm{kpc}$ is the distance from Sgr A* to Earth \cite{abuter2019geometric}.
If one assumes that the scales of the CMZ are $-50\,\mathrm{pc}<z<50\,\mathrm{pc}$ and $r<200\,\mathrm{pc}$, then the fluxes of the gamma rays and neutrinos are
\begin{eqnarray}
    F_{\gamma, \nu}&\sim&\frac{cn_\mathrm{H}}{16\pi^2(8\,\mathrm{kpc})^2}Y_{\gamma, \nu}\left(E_{\gamma, \nu}\right)\nonumber\\
    & &\times\int_{0}^{2\pi}\int_{-50\,\mathrm{pc}}^{50\,\mathrm{pc}}\int_{0}^{200\,\mathrm{pc}}\frac{1}{r}\, rdrdzd\theta\nonumber\\
    &=&\frac{cn_\mathrm{H}}{25600\pi}Y_{\gamma, \nu}\left(E_{\gamma, \nu}\right),
\end{eqnarray}
where
\begin{eqnarray}
\label{Ygn}
    Y_{\gamma, \nu}\left(E_{\gamma, \nu}\right)=\qquad\qquad\qquad\qquad\qquad\qquad\nonumber\\
    \int_{E_{\gamma, \nu}}^{\infty}\sigma_{\mathrm{inel}}\left(E_p\right)\frac{Q\left(E_p\right)}{D\left(E_p\right)}F_{\gamma, \nu}\left(\frac{E_{\gamma, \nu}}{E_p}, E_p\right)\frac{dE_p}{E_p}.
\end{eqnarray}
We show in the Fig. \ref{gammadif} the fluxes of the gamma rays and the neutrinos if the gas in the CMZ has the typical density of $n_\mathrm{H}\sim100\,\mathrm{cm}^{-3}$ \cite{dahmen1998molecular} and diffusion coefficient $D\left(E_p\right)\sim3\times{10}^{28}\left(E_p/\mathrm{GeV}\right)^{1/3}\,\mathrm{cm}^{2}\,\mathrm{s}^{-1}$ along with the flux of the gamma ray data from HESS and MAGIC collaboration  \cite{collaboration2016acceleration,acciari2020magic}. The flux ${E_{\gamma}}^2F_{\gamma}$ has a peak value $1.1\times10^{-13}\,\mathrm{TeV\,cm^{-2}\,s^{-1}}$ at $61\,\mathrm{TeV}$ for $B_0=100\,\mathrm{G}$ and $4.3\times10^{-14}\,\mathrm{TeV\,cm^{-2}\,s^{-1}}$ at $19\,\mathrm{TeV}$ for $B_0=30\,\mathrm{G}$.

\begin{figure}
\includegraphics[width=8.66cm]{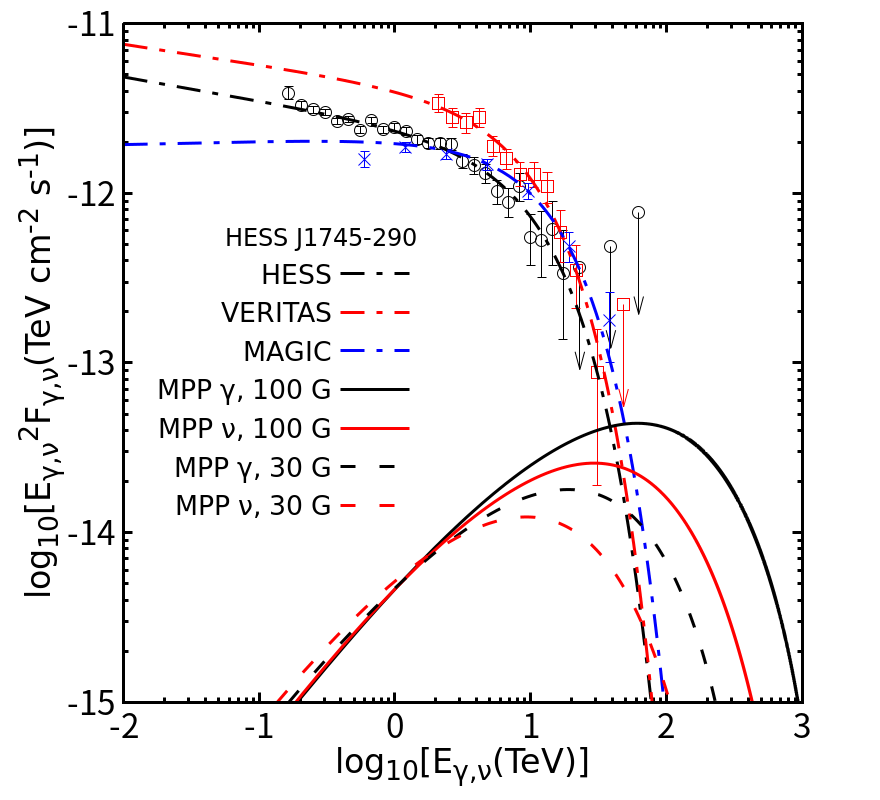}
\caption{Observation data of HESS J1745-290 (black open circles from HESS \cite{collaboration2016acceleration}, blue crosses from MAGIC \cite{acciari2020magic} and red rectangles from VERITAS \cite{adams2021veritas}), and the fluxes of gamma rays and neutrinos from protons which are accelerated by the MPP in Sgr A*.}
\label{gamma1745}
\end{figure}

From the information on the gas density at 0.1° circular region (inner $\sim10\,\mathrm{pc}$ region), one can deduce the contribution of the MPP on HESS J1745-290. At a few parsec scales around Sgr A*, the gas density shows a distribution of $\sim10^{3}\,\mathrm{cm}^{-3}$ on average, and there is a dense region at $1-3\,\mathrm{pc}$ \cite{ferriere2007spatial,ferriere2012interstellar}. The gas density averaged over solid angle is about $8\times10^{4}\,\mathrm{cm}^{-3}$ at about $1-3\,\mathrm{pc}$ and about $2\times10^{3}\,\mathrm{cm}^{-3}$ at about $0-1\,\mathrm{pc}$ and $3-10\,\mathrm{pc}$ \cite{linden2012morphology}. Using the gas density and the Eq. \eqref{fgn}, we determine the gamma nray flux from accelerated protons by the MPP, ${E_{\gamma}}^2F_{\gamma}$. It has peak values $4.4\times10^{-14}\,\mathrm{TeV\,cm^{-2}\,s^{-1}}$ at $61\,\mathrm{TeV}$ for $B_0=100\,\mathrm{G}$ and $1.8\times10^{-14}\,\mathrm{TeV\,cm^{-2}\,s^{-1}}$ at $19\,\mathrm{TeV}$ for $B_0=30\,\mathrm{G}$ (Fig. \ref{gamma1745}).

Figures \ref{gammadif} and \ref{gamma1745} show that the flux from MPP can have a significant effect on the gamma ray flux of CMZ and HESS J1745-290 at $E_{\gamma}\gtrsim10\,\mathrm{TeV}$. The flux from MPP in the case of $100\,\mathrm{G}$ ($30\,\mathrm{G}$) is about $3.5\%$ ($2.3\%$) of the observed gamma ray flux of CMZ at about $10\,\mathrm{TeV}$. The flux from MPP in the case of $100\,\mathrm{G}$ ($30\,\mathrm{G}$) is about $1.0-2.4\%$ ($0.7-1.6\%$) of observed gamma ray flux of HESS J1745-290 at about $10\,\mathrm{TeV}$.

The flux from MPP surpasses the broken power law based on gamma ray flux of CMZ from MAGIC (blue dot-dashed line) at about $63\,\mathrm{TeV}$ in the case of $100\,\mathrm{G}$ and surpasses the broken power law at about $95\,\mathrm{TeV}$ in the case of $30\,\mathrm{G}$. The flux from MPP is lower than the power law based on gamma ray flux of CMZ from HESS (black dot-dashed line) at every energy region. The flux from MPP surpasses the broken power law based on gamma ray flux of HESS J1745-290 from HESS (black dot dashed line), MAGIC (blue dot dashed line), and VERITAS (red dot dashed line) data at about $49\,\mathrm{TeV}$ in the case of $100\,\mathrm{G}$ and surpasses the broken power law at about $67\,\mathrm{TeV}$ in the case of $30\,\mathrm{G}$. These results show that detailed future observation of gamma ray flux of CMZ and HESS J1745-290 at $\lesssim100\,\mathrm{TeV}$ may provide the existence of MPP or limit of parameters for MPP.

\section{Accelerated proton cosmic rays flux}
In the range of $E\gtrsim10^{10}\,\mathrm{eV}$, cosmic rays are not affected by solar modulation and the cosmic ray spectrum shows a power law spectrum with knee structures at few $10^{15}\,\mathrm{eV}$ and at $10^{17}-10^{18}\,\mathrm{eV}$. Random Galactic magnetic field effectively confines cosmic rays in the Galaxy and makes an anisotropic flux depending on the location of the source to an isotropic flux by scattering. The diffusion approximates such transport of cosmic rays. The mean magnetic field strength of the Galaxy is about $3\,\mathrm{\mu G}$ and the scale of the random magnetic field is about $100\,\mathrm{pc}$ \cite{ptuskin2006cosmic}. So one can use the diffusion equation to estimate the transport of the accelerated protons from MPP at Sgr A* to Earth if the gyroradius of the proton is less than $100\,\mathrm{pc}$. Gyroradius is given by
\begin{equation}
\label{cr1}
    r_{g}=\frac{p}{|q|B}=\frac{E}{|q|Bc}\simeq0.3\left(\frac{E}{10^{15}\,\mathrm{eV}}\right)\,\mathrm{pc}.
\end{equation}
The peak energy of the accelerated proton from Sgr A* in our model is about $0.5\times10^{15}\,\mathrm{eV}$ for $100\,\mathrm{G}$ magnetic field around Sgr A*, much greater than the proton mass of $\sim10^{9}\,\mathrm{eV}$, and one can use $p=E/c$ in Eq. \eqref{cr1}. The gyroradius of the proton with $0.5\times10^{15}\,\mathrm{eV}$ is about $0.1\,\mathrm{pc}$, and diffusion approximation is valid.

When the number of cosmic rays in the Galaxy is stationary and there are no energy loss and reacceleration, the diffusion equation is
\begin{equation}
\label{cr2}
    \boldsymbol{\bigtriangledown}\cdot\boldsymbol{j}=q(\boldsymbol{x}),\quad j_{i}=-D_{ij}(\boldsymbol{x})\triangledown_{j}n,
\end{equation}
where $q(\boldsymbol{x})$ is the time independent source term, $\boldsymbol{j}$ the cosmic ray current, and $D_{ij}$ the diffusive coefficient tensor \cite{ptuskin1993diffusion,candia2003turbulent}. We assume that the diffusive coefficient tensor only depends on the energy of the cosmic ray to simplify Eq. \eqref{cr2}. The diffusion coefficient can be scaled by the cosmic ray rigidity:
\begin{equation}
\label{cr3}
    D=D_{0}\beta R^{\delta},\quad \beta=\frac{v}{c},\quad R=\frac{pc}{ze},
\end{equation}
where $R$, $z$, and $e$ are rigidity, atomic number, and elementary charge, respectively. Most models estimate $\delta=1/3$ and this is consistent with the Kolmogorov spectrum \cite{ptuskin2006cosmic,evoli2018origin}. Thus, we use
\begin{equation}
\label{cr4}
    D_{K}=3\times10^{28}\beta\left(R/\mathrm{GV}\right)^{1/3}\,\mathrm{cm^{2} s^{-1}},
\end{equation}
where $\mathrm{GV}$ is gigavolt. Sgr A* is effectively a delta function source in the Galaxy and Eq. \eqref{cr2} has the same form as the central force. Therefore,
\begin{equation}
\label{cr5}
    \frac{dn}{dE_{p}}=\frac{Q_{MPP}}{4\pi Dr_s},\quad q(\boldsymbol{x})=Q_{MPP}\delta(\boldsymbol{x}),
\end{equation}
where $r_{s}$ is the distance from the source to Earth.
The main structure of the Galaxy is a disklike, not spherical. Taillet and Maurin~\cite{taillet2003spatial} use equation \eqref{cr2} with to obtain top-down boundaries with diffusive volume. The particle distribution model:
\begin{equation}
\label{cr6}
    \frac{dn}{dE_{p}}=\frac{Q_{MPP}}{4\pi Dr_{s}}\times2\sqrt{\frac{r_{s}}{L}}e^{-\pi r_{s}/2L},
\end{equation}
where $L$ is the height of the Galactic halo (diffusive volume). Equation \eqref{cr6} does not consider the side boundary of the Galaxy. However, the equation fits well in $L\lesssim10\,\mathrm{kpc}$ with the model that considers all boundaries of the Galaxy. Analyse of various cosmic rays observed on Earth, Refs.~\cite{evoli2018origin,beck2001galactic,orlando2013galactic} estimate $L=4\,\mathrm{kpc}$, $L\simeq6\,\mathrm{kpc}$, and $L=10\,\mathrm{kpc}$, respectively. These results ensure that we just use Eq. \eqref{cr6}. We use $L=4\,\mathrm{kpc}$ with Eq. \eqref{cr6} from Ref. \cite{ptuskin2006cosmic} that presented $L=4\,\mathrm{kpc}$ with static diffusion model without the Galactic wind. We assume the diffused protons flow to be isotropic, and the accelerated proton flux to Earth from Sgr A* is
\begin{equation}
\label{cr7}
    F_{MPP}(E_p)=\frac{Q_{MPP}c}{16\pi Dr_s}\times2\sqrt{\frac{r_s}{L}}e^{-\pi r_s/2L},
\end{equation}
where $r_s$ is the distance from Sgr A* to Earth, chosen to be $8\,\mathrm{kpc}$ \cite{abuter2019geometric}. 

\begin{figure}
\includegraphics[width=8.66cm]{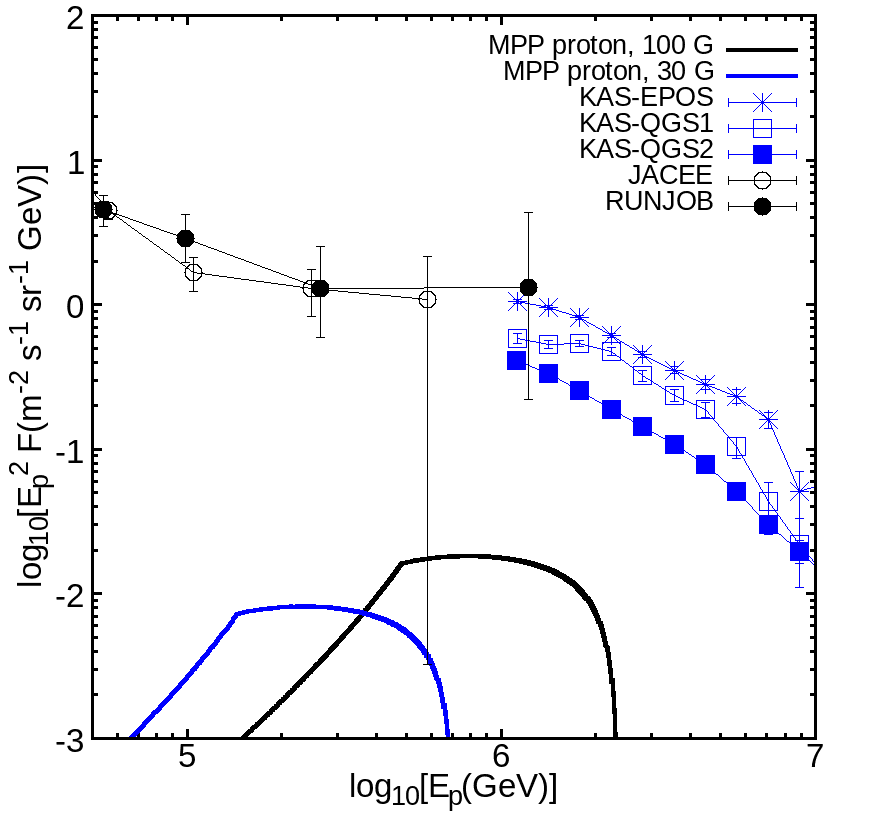}
\caption{The accelerated proton flux (black solid and blue solid line) from Sgr A* estimated by Eqs. \eqref{Qmpp} and \eqref{cr7} and proton cosmic ray flux from JACEE (black, open circle) \cite{asakimori1998cosmic} and RUNJOB (black, filled circle) \cite{balloon2001composition} balloon experiment and KASCADE experiment with QGSJET01 (blue, filled rectangle), QGSJET\uppercase\expandafter{\romannumeral2} (blue, open triangle), and EPOS1.99 (blue, cross) \cite{finger2011reconstruction}.}
\label{cos}
\end{figure}

Figure \ref{cos} shows that proton cosmic ray flux from JACEE (black, open circle) \cite{asakimori1998cosmic} and RUNJOB (black, filled circle) \cite{balloon2001composition} balloon experiment data,  KASCADE (blue rectangle and cross) \cite{finger2011reconstruction} experiment data and the accelerated proton flux from Sgr A* (solid lines) estimated by Eqs. \eqref{Qmpp} and \eqref{cr7}. Compared to JACEE data, the accelerated proton flux is $0.2\%$ (${E_p}^2F_{MPP}=2.9\times10^{-3}\,\mathrm{m^{-2}\,s^{-1}\,sr^{-1}\,GeV}$) in the case of $B_0=100\,\mathrm{G}$ (black solid line) and the accelerated proton flux is $0.6\%$ (${E_p}^2F_{MPP}=8.1\times10^{-3}\,\mathrm{m^{-2}\,s^{-1}\,sr^{-1}\,GeV}$) in the case of $B_0=30\,\mathrm{G}$ (blue solid line) at $0.25\,\mathrm{PeV}$. The accelerated proton flux is about 1.6\%, 2.9\%, and 4.1\% of cosmic ray proton flux with KASCADE EPOS1.99 (blue, open rectangle), QGSJET01 (black, filled rectangle) and QGSJET\uppercase\expandafter{\romannumeral2} (orange, open triangle), respectively, at $E_{p}=1.1\,\mathrm{PeV}$ (${E_p}^2F_{MPP}=9.1\times10^{-2}\,\mathrm{m^{-2}\,s^{-1}\,sr^{-1}\,GeV}$). 

\section{Conclusion and discussion}
MPP considers the electromagnetic interaction, unlike the normal Penrose process. These feature not only removes the limit of the normal Penrose process but can also accelerate charged particle to high energy. We calculate the production rate of the high-energy protons accelerated by MPP at Sgr A*. We also estimate the flux of gamma rays and the flux of the accelerated protons diffused from Sgr A* to Earth. 

We model the accretion flow onto Sgr A* with the simple ADAF model that is hot enough to produce neutrons. We then estimate how many neutrons are produced within the ADAF and transported into the accelerated zone. Neutrons decay into protons, that are subsequently accelerated along with rotational axis of the black hole in MPP by electric potential difference described by Wald solution. The peak production rate of high-energy accelerated protons is $5.5\times10^{18}\,\mathrm{eV^{-1}\,s^{-1}}$ at $0.48\,\mathrm{TeV}$ in the case of uniform magnetic field strength around Sgr A* of $B_0=100\,\mathrm{G}$ and $1.8\times10^{19}\,\mathrm{eV^{-1}\,s^{-1}}$ at $0.15\,\mathrm{TeV}$ in the case of $B_0=30\,\mathrm{G}$.

Radiation fluxes of synchrotron, inverse Compton and bremsstrahlung of the high-energy protons do not affect the current radiation flux of Sgr A*. On the other hand, the flux of VHE gamma rays from p-p interaction of the high-energy protons against GC hydrogen significantly affects the flux of VHE gamma rays of CMZ and HESS J1745-290 at $\ge 10\,\mathrm{TeV}$. The peak flux ${E_\gamma}^2F_\gamma$ from MPP on CMZ scale are $1.3\times10^{-13}\,\mathrm{TeV\,cm^{-2}\,s^{-1}}$ at $61\,\mathrm{TeV}$ for $B_0=100\,\mathrm{G}$ and $4.2\times10^{-14}\,\mathrm{TeV\,cm^{-2}\,s^{-1}}$ at $19\,\mathrm{TeV}$ for $B_0=30\,\mathrm{G}$. The peak flux ${E_\gamma}^2F_\gamma$ from MPP on HESS J1745-290 scale are $4.4\times10^{-14}\,\mathrm{TeV\,cm^{-2}\,s^{-1}}$ at $61\,\mathrm{TeV}$ for $B_0=100\,\mathrm{G}$ and $1.8\times10^{-14}\,\mathrm{TeV\,cm^{-2}\,s^{-1}}$ at $19\,\mathrm{TeV}$ for $B_0=30\,\mathrm{G}$. At $\geq 50-100\,\mathrm{TeV}$, The flux ${E_\gamma}^2F_\gamma$ from MPP surpasses the expected flux from the broken power law fit based on observed data. This result implies that precise future observation at $\leq100\,\mathrm{TeV}$ will provide the existence of MPP or a limit of parameters of MPP.

The gyroradius of high-energy protons is much smaller than the scale of the Galaxy. We estimate flux at Earth with the diffusion equation that is considered for stationary high-energy proton production. MPP proton flux is about $1.6-4.1\%$ of cosmic ray proton flux from KASCADE experiment at about $1\,\mathrm{PeV}$.

The essential conditions for our model to operate are whether the electric field from Wald solution is not screened, the ability to produce neutrons and the timescale of the neutron inflow. We can ask if the same process can operate in other SMBHs, such as M87. Eq. \eqref{thresB} is the threshold magnetic field strength to produce the electron-positron pair cascade under the force-free condition. The magnetic field strength in the vicinity of the black hole in M87 is $1-30\,\mathrm{G}$ \cite{akiyama2021first}, which is comparable to the threshold magnetic field strength. Previous GRMHD simulations studies of the pair production in M87 also present that the pair density at the jet spine in the vicinity of the black hole is much higher than the Goldreich-Julian density \cite{wong2021pair, moscibrodzka2016general}. This pair density makes our model difficult to operate in M87.

The neutron production rate scales with the mass accretion rate and the mass of the black hole. From the self-similar ADAF solutions, the physical gas density of the accretion flow is $\rho\propto\dot{m}m^{-1}$. The reaction rate of neutron production nuclear reactions is $R_{reaction}\propto{\dot{m}}^2m^{-2}$ by Eq. \eqref{reaction1}. Since the total number of produced neutrons per unit time $\dot{N}$ is proportional to the volume, $\dot{N}\propto{\dot{m}}^2m$, which shows that heavy SMBH is more efficient in neutron production for the same value of $\dot{m}$.

However, heavy SMBHs like M87 are critically disadvantaged due to the longer timescale of the neutron transport from the accretion flow to the acceleration zone. Assuming that the transport distance is approximately of the order of Schwarzschild radius and the speed of the neutron is approximately the speed of light, the number of neutrons that arrive into the acceleration zone per unit time is $\dot{N}_{in}\propto{\dot{m}}^2m\exp(-2GM_{\odot}m/c^{3}\tau_n)$, where $\tau_n$ is the mean lifetime of the neutron. 

Since this exponential factor will be dominant for large $m$, even when the acceleration zone and the accretion flow are favorably configured, the accretion flow can supply neutrons only when $m$ is not much greater than $9\times10^{7}$. Therefore, the acceleration model considered in this work will operate only in SMBHs with mass not much greater than $\sim10^8\,M_\odot$, and M87 is much heavier than this critical mass.

\begin{acknowledgments}
The authors thank Prof. Remo Ruffini for insightful discussion and the anonymous referee for many useful and insightful comments that have improved this paper. This work was supported by the Basic Science Research Program through the National Research Foundation of Korea Grant No. 2019R1I1A3A02062242, No. 2018R1A6A1A06024970, and RS-2023-00240212.
\end{acknowledgments}

\nocite{*}

\bibliography{paper}

\providecommand{\noopsort}[1]{}\providecommand{\singleletter}[1]{#1}%
\begin{thebibliography}{96}%
\makeatletter
\providecommand \@ifxundefined [1]{%
 \@ifx{#1\undefined}
}%
\providecommand \@ifnum [1]{%
 \ifnum #1\expandafter \@firstoftwo
 \else \expandafter \@secondoftwo
 \fi
}%
\providecommand \@ifx [1]{%
 \ifx #1\expandafter \@firstoftwo
 \else \expandafter \@secondoftwo
 \fi
}%
\providecommand \natexlab [1]{#1}%
\providecommand \enquote  [1]{``#1''}%
\providecommand \bibnamefont  [1]{#1}%
\providecommand \bibfnamefont [1]{#1}%
\providecommand \citenamefont [1]{#1}%
\providecommand \href@noop [0]{\@secondoftwo}%
\providecommand \href [0]{\begingroup \@sanitize@url \@href}%
\providecommand \@href[1]{\@@startlink{#1}\@@href}%
\providecommand \@@href[1]{\endgroup#1\@@endlink}%
\providecommand \@sanitize@url [0]{\catcode `\\12\catcode `\$12\catcode
  `\&12\catcode `\#12\catcode `\^12\catcode `\_12\catcode `\%12\relax}%
\providecommand \@@startlink[1]{}%
\providecommand \@@endlink[0]{}%
\providecommand \url  [0]{\begingroup\@sanitize@url \@url }%
\providecommand \@url [1]{\endgroup\@href {#1}{\urlprefix }}%
\providecommand \urlprefix  [0]{URL }%
\providecommand \Eprint [0]{\href }%
\providecommand \doibase [0]{https://doi.org/}%
\providecommand \selectlanguage [0]{\@gobble}%
\providecommand \bibinfo  [0]{\@secondoftwo}%
\providecommand \bibfield  [0]{\@secondoftwo}%
\providecommand \translation [1]{[#1]}%
\providecommand \BibitemOpen [0]{}%
\providecommand \bibitemStop [0]{}%
\providecommand \bibitemNoStop [0]{.\EOS\space}%
\providecommand \EOS [0]{\spacefactor3000\relax}%
\providecommand \BibitemShut  [1]{\csname bibitem#1\endcsname}%
\let\auto@bib@innerbib\@empty
\bibitem [{\citenamefont {Christodoulou}(1970)}]{christodoulou1970reversible}%
  \BibitemOpen
  \bibfield  {author} {\bibinfo {author} {\bibfnamefont {D.}~\bibnamefont
  {Christodoulou}},\ }\href {https://doi.org/10.1103/PhysRevLett.25.1596}
  {\bibfield  {journal} {\bibinfo  {journal} {Phys. Rev. Lett.}\ }\textbf
  {\bibinfo {volume} {25}},\ \bibinfo {pages} {1596} (\bibinfo {year}
  {1970})}\BibitemShut {NoStop}%
\bibitem [{\citenamefont {Christodoulou}\ and\ \citenamefont
  {Ruffini}(1971)}]{christodoulou1971reversible}%
  \BibitemOpen
  \bibfield  {author} {\bibinfo {author} {\bibfnamefont {D.}~\bibnamefont
  {Christodoulou}}\ and\ \bibinfo {author} {\bibfnamefont {R.}~\bibnamefont
  {Ruffini}},\ }\href {https://doi.org/10.1103/PhysRevD.4.3552} {\bibfield
  {journal} {\bibinfo  {journal} {Phys. Rev. D}\ }\textbf {\bibinfo {volume}
  {4}},\ \bibinfo {pages} {3552} (\bibinfo {year} {1971})}\BibitemShut
  {NoStop}%
\bibitem [{\citenamefont {Hawking}(1971)}]{hawking1971gravitational}%
  \BibitemOpen
  \bibfield  {author} {\bibinfo {author} {\bibfnamefont {S.~W.}\ \bibnamefont
  {Hawking}},\ }\href {https://doi.org/10.1103/PhysRevLett.26.1344} {\bibfield
  {journal} {\bibinfo  {journal} {Phys. Rev. Lett.}\ }\textbf {\bibinfo
  {volume} {26}},\ \bibinfo {pages} {1344} (\bibinfo {year}
  {1971})}\BibitemShut {NoStop}%
\bibitem [{\citenamefont {Penrose}(1969)}]{penrose1969gravitational}%
  \BibitemOpen
  \bibfield  {author} {\bibinfo {author} {\bibfnamefont {R.}~\bibnamefont
  {Penrose}},\ }\href
  {https://link.springer.com/article/10.1023/A:1016578408204} {\bibfield
  {journal} {\bibinfo  {journal} {Nuovo Cimento Riv. Ser.}\ }\textbf {\bibinfo
  {volume} {1}},\ \bibinfo {pages} {252} (\bibinfo {year} {1969})}\BibitemShut
  {NoStop}%
\bibitem [{\citenamefont {Bardeen}\ \emph {et~al.}(1972)\citenamefont
  {Bardeen}, \citenamefont {Press},\ and\ \citenamefont
  {Teukolsky}}]{bardeen1972rotating}%
  \BibitemOpen
  \bibfield  {author} {\bibinfo {author} {\bibfnamefont {J.~M.}\ \bibnamefont
  {Bardeen}}, \bibinfo {author} {\bibfnamefont {W.~H.}\ \bibnamefont {Press}},\
  and\ \bibinfo {author} {\bibfnamefont {S.~A.}\ \bibnamefont {Teukolsky}},\
  }\href {https://doi.org/10.1086/151796} {\bibfield  {journal} {\bibinfo
  {journal} {Astrophys. J.}\ }\textbf {\bibinfo {volume} {178}},\ \bibinfo
  {pages} {347} (\bibinfo {year} {1972})}\BibitemShut {NoStop}%
\bibitem [{\citenamefont {Wald}(1974{\natexlab{a}})}]{wald1974energy}%
  \BibitemOpen
  \bibfield  {author} {\bibinfo {author} {\bibfnamefont {R.~M.}\ \bibnamefont
  {Wald}},\ }\href {https://doi.org/10.1086/152959} {\bibfield  {journal}
  {\bibinfo  {journal} {Astrophys. J.}\ }\textbf {\bibinfo {volume} {191}},\
  \bibinfo {pages} {231} (\bibinfo {year} {1974}{\natexlab{a}})}\BibitemShut
  {NoStop}%
\bibitem [{\citenamefont {Wagh}\ \emph {et~al.}(1985)\citenamefont {Wagh},
  \citenamefont {Dhurandhar},\ and\ \citenamefont {Dadhich}}]{wagh1985revival}%
  \BibitemOpen
  \bibfield  {author} {\bibinfo {author} {\bibfnamefont {S.}~\bibnamefont
  {Wagh}}, \bibinfo {author} {\bibfnamefont {S.}~\bibnamefont {Dhurandhar}},\
  and\ \bibinfo {author} {\bibfnamefont {N.}~\bibnamefont {Dadhich}},\ }\href
  {https://doi.org/10.1086/162952} {\bibfield  {journal} {\bibinfo  {journal}
  {Astrophys. J.}\ }\textbf {\bibinfo {volume} {290}},\ \bibinfo {pages} {12}
  (\bibinfo {year} {1985})}\BibitemShut {NoStop}%
\bibitem [{\citenamefont {Parthasarathy}\ \emph {et~al.}(1986)\citenamefont
  {Parthasarathy}, \citenamefont {Wagh}, \citenamefont {Dhurandhar},\ and\
  \citenamefont {Dadhich}}]{parthasarathy1986high}%
  \BibitemOpen
  \bibfield  {author} {\bibinfo {author} {\bibfnamefont {S.}~\bibnamefont
  {Parthasarathy}}, \bibinfo {author} {\bibfnamefont {S.}~\bibnamefont {Wagh}},
  \bibinfo {author} {\bibfnamefont {S.}~\bibnamefont {Dhurandhar}},\ and\
  \bibinfo {author} {\bibfnamefont {N.}~\bibnamefont {Dadhich}},\ }\href
  {https://doi.org/10.1086/164390} {\bibfield  {journal} {\bibinfo  {journal}
  {Astrophys. J.}\ }\textbf {\bibinfo {volume} {307}},\ \bibinfo {pages} {38}
  (\bibinfo {year} {1986})}\BibitemShut {NoStop}%
\bibitem [{\citenamefont {Narayan}\ \emph {et~al.}(2014)\citenamefont
  {Narayan}, \citenamefont {McClintock},\ and\ \citenamefont
  {Tchekhovskoy}}]{narayan2014energy}%
  \BibitemOpen
  \bibfield  {author} {\bibinfo {author} {\bibfnamefont {R.}~\bibnamefont
  {Narayan}}, \bibinfo {author} {\bibfnamefont {J.~E.}\ \bibnamefont
  {McClintock}},\ and\ \bibinfo {author} {\bibfnamefont {A.}~\bibnamefont
  {Tchekhovskoy}},\ }in\ \href {https://doi.org/10.1007/978-3-319-06349-2_25}
  {\emph {\bibinfo {booktitle} {General Relativity, Cosmology and
  Astrophysics}}}\ (\bibinfo  {publisher} {Springer},\ \bibinfo {address} {New
  York},\ \bibinfo {year} {2014})\ p.\ \bibinfo {pages} {523}\BibitemShut
  {NoStop}%
\bibitem [{\citenamefont {Blandford}\ and\ \citenamefont
  {Znajek}(1977)}]{blandford1977electromagnetic}%
  \BibitemOpen
  \bibfield  {author} {\bibinfo {author} {\bibfnamefont {R.~D.}\ \bibnamefont
  {Blandford}}\ and\ \bibinfo {author} {\bibfnamefont {R.~L.}\ \bibnamefont
  {Znajek}},\ }\href {https://doi.org/10.1093/mnras/179.3.433} {\bibfield
  {journal} {\bibinfo  {journal} {Mon. Not. R. Astron. Soc.}\ }\textbf
  {\bibinfo {volume} {179}},\ \bibinfo {pages} {433} (\bibinfo {year}
  {1977})}\BibitemShut {NoStop}%
\bibitem [{\citenamefont {Abuter}\ \emph {et~al.}(2019)\citenamefont {Abuter}
  \emph {et~al.}}]{abuter2019geometric}%
  \BibitemOpen
  \bibfield  {author} {\bibinfo {author} {\bibfnamefont {R.}~\bibnamefont
  {Abuter}} \emph {et~al.} (\bibinfo {collaboration} {The GRAVITY
  Collaboration}),\ }\href {https://doi.org/10.1051/0004-6361/201935656}
  {\bibfield  {journal} {\bibinfo  {journal} {Astron. Astrophys.}\ }\textbf
  {\bibinfo {volume} {625}},\ \bibinfo {pages} {L10} (\bibinfo {year}
  {2019})}\BibitemShut {NoStop}%
\bibitem [{\citenamefont {Dadhich}\ \emph {et~al.}(2018)\citenamefont
  {Dadhich}, \citenamefont {Tursunov}, \citenamefont {Ahmedov},\ and\
  \citenamefont {Stuchl{\'\i}k}}]{dadhich2018distinguishing}%
  \BibitemOpen
  \bibfield  {author} {\bibinfo {author} {\bibfnamefont {N.}~\bibnamefont
  {Dadhich}}, \bibinfo {author} {\bibfnamefont {A.}~\bibnamefont {Tursunov}},
  \bibinfo {author} {\bibfnamefont {B.}~\bibnamefont {Ahmedov}},\ and\ \bibinfo
  {author} {\bibfnamefont {Z.}~\bibnamefont {Stuchl{\'\i}k}},\ }\href
  {https://doi.org/10.1093/mnrasl/sly073} {\bibfield  {journal} {\bibinfo
  {journal} {Mon. Not. R. Astron. Soc. Lett.}\ }\textbf {\bibinfo {volume}
  {478}},\ \bibinfo {pages} {L89} (\bibinfo {year} {2018})}\BibitemShut
  {NoStop}%
\bibitem [{\citenamefont {Wald}(1974{\natexlab{b}})}]{wald1974black}%
  \BibitemOpen
  \bibfield  {author} {\bibinfo {author} {\bibfnamefont {R.~M.}\ \bibnamefont
  {Wald}},\ }\href {https://doi.org/10.1103/PhysRevD.10.1680} {\bibfield
  {journal} {\bibinfo  {journal} {Phys. Rev. D}\ }\textbf {\bibinfo {volume}
  {10}},\ \bibinfo {pages} {1680} (\bibinfo {year}
  {1974}{\natexlab{b}})}\BibitemShut {NoStop}%
\bibitem [{\citenamefont {Stuchl{\'\i}k}\ and\ \citenamefont
  {Kolo{\v{s}}}(2016)}]{stuchlik2016acceleration}%
  \BibitemOpen
  \bibfield  {author} {\bibinfo {author} {\bibfnamefont {Z.}~\bibnamefont
  {Stuchl{\'\i}k}}\ and\ \bibinfo {author} {\bibfnamefont {M.}~\bibnamefont
  {Kolo{\v{s}}}},\ }\href {https://doi.org/10.1140/epjc/s10052-015-3862-2}
  {\bibfield  {journal} {\bibinfo  {journal} {Eur. Phys. J. C}\ }\textbf
  {\bibinfo {volume} {76}},\ \bibinfo {pages} {1} (\bibinfo {year}
  {2016})}\BibitemShut {NoStop}%
\bibitem [{\citenamefont {Tursunov}\ \emph {et~al.}(2016)\citenamefont
  {Tursunov}, \citenamefont {Stuchl{\'\i}k},\ and\ \citenamefont
  {Kolo{\v{s}}}}]{tursunov2016circular}%
  \BibitemOpen
  \bibfield  {author} {\bibinfo {author} {\bibfnamefont {A.}~\bibnamefont
  {Tursunov}}, \bibinfo {author} {\bibfnamefont {Z.}~\bibnamefont
  {Stuchl{\'\i}k}},\ and\ \bibinfo {author} {\bibfnamefont {M.}~\bibnamefont
  {Kolo{\v{s}}}},\ }\href {https://doi.org/10.1103/PhysRevD.93.084012}
  {\bibfield  {journal} {\bibinfo  {journal} {Phys. Rev. D}\ }\textbf {\bibinfo
  {volume} {93}},\ \bibinfo {pages} {084012} (\bibinfo {year}
  {2016})}\BibitemShut {NoStop}%
\bibitem [{\citenamefont {Kolo{\v{s}}}\ \emph {et~al.}(2017)\citenamefont
  {Kolo{\v{s}}}, \citenamefont {Tursunov},\ and\ \citenamefont
  {Stuchl{\'\i}k}}]{kolovs2017possible}%
  \BibitemOpen
  \bibfield  {author} {\bibinfo {author} {\bibfnamefont {M.}~\bibnamefont
  {Kolo{\v{s}}}}, \bibinfo {author} {\bibfnamefont {A.}~\bibnamefont
  {Tursunov}},\ and\ \bibinfo {author} {\bibfnamefont {Z.}~\bibnamefont
  {Stuchl{\'\i}k}},\ }\href {https://doi.org/10.1140/epjc/s10052-017-5431-3}
  {\bibfield  {journal} {\bibinfo  {journal} {Eur. Phys. J. C}\ }\textbf
  {\bibinfo {volume} {77}},\ \bibinfo {pages} {1} (\bibinfo {year}
  {2017})}\BibitemShut {NoStop}%
\bibitem [{\citenamefont {Ruffini}\ \emph {et~al.}(2019)\citenamefont {Ruffini}
  \emph {et~al.}}]{ruffini2019gev}%
  \BibitemOpen
  \bibfield  {author} {\bibinfo {author} {\bibfnamefont {R.}~\bibnamefont
  {Ruffini}} \emph {et~al.},\ }\href {https://doi.org/10.3847/1538-4357/ab4ce6}
  {\bibfield  {journal} {\bibinfo  {journal} {Astrophys. J.}\ }\textbf
  {\bibinfo {volume} {886}},\ \bibinfo {pages} {82} (\bibinfo {year}
  {2019})}\BibitemShut {NoStop}%
\bibitem [{\citenamefont {Rueda}\ and\ \citenamefont
  {Ruffini}(2020)}]{rueda2020blackholic}%
  \BibitemOpen
  \bibfield  {author} {\bibinfo {author} {\bibfnamefont {J.}~\bibnamefont
  {Rueda}}\ and\ \bibinfo {author} {\bibfnamefont {R.}~\bibnamefont
  {Ruffini}},\ }\href {https://doi.org/10.1140/epjc/s10052-020-7868-z}
  {\bibfield  {journal} {\bibinfo  {journal} {Eur. Phys. J. C}\ }\textbf
  {\bibinfo {volume} {80}},\ \bibinfo {pages} {1} (\bibinfo {year}
  {2020})}\BibitemShut {NoStop}%
\bibitem [{\citenamefont {Tursunov}\ \emph {et~al.}(2020)\citenamefont
  {Tursunov}, \citenamefont {Stuchl{\'\i}k}, \citenamefont {Kolo{\v{s}}},
  \citenamefont {Dadhich},\ and\ \citenamefont
  {Ahmedov}}]{tursunov2020supermassive}%
  \BibitemOpen
  \bibfield  {author} {\bibinfo {author} {\bibfnamefont {A.}~\bibnamefont
  {Tursunov}}, \bibinfo {author} {\bibfnamefont {Z.}~\bibnamefont
  {Stuchl{\'\i}k}}, \bibinfo {author} {\bibfnamefont {M.}~\bibnamefont
  {Kolo{\v{s}}}}, \bibinfo {author} {\bibfnamefont {N.}~\bibnamefont
  {Dadhich}},\ and\ \bibinfo {author} {\bibfnamefont {B.}~\bibnamefont
  {Ahmedov}},\ }\href {https://doi.org/10.3847/1538-4357/ab8ae9} {\bibfield
  {journal} {\bibinfo  {journal} {Astrophys. J.}\ }\textbf {\bibinfo {volume}
  {895}},\ \bibinfo {pages} {14} (\bibinfo {year} {2020})}\BibitemShut
  {NoStop}%
\bibitem [{\citenamefont {Moradi}\ \emph {et~al.}(2021)\citenamefont {Moradi},
  \citenamefont {Rueda}, \citenamefont {Ruffini},\ and\ \citenamefont
  {Wang}}]{moradi2021newborn}%
  \BibitemOpen
  \bibfield  {author} {\bibinfo {author} {\bibfnamefont {R.}~\bibnamefont
  {Moradi}}, \bibinfo {author} {\bibfnamefont {J.}~\bibnamefont {Rueda}},
  \bibinfo {author} {\bibfnamefont {R.}~\bibnamefont {Ruffini}},\ and\ \bibinfo
  {author} {\bibfnamefont {Y.}~\bibnamefont {Wang}},\ }\href
  {https://doi.org/10.1051/0004-6361/201937135} {\bibfield  {journal} {\bibinfo
   {journal} {Astron. Astrophys.}\ }\textbf {\bibinfo {volume} {649}},\
  \bibinfo {pages} {A75} (\bibinfo {year} {2021})}\BibitemShut {NoStop}%
\bibitem [{\citenamefont {Rueda}\ \emph {et~al.}(2022)\citenamefont {Rueda},
  \citenamefont {Ruffini},\ and\ \citenamefont
  {Kerr}}]{rueda2022gravitomagnetic}%
  \BibitemOpen
  \bibfield  {author} {\bibinfo {author} {\bibfnamefont {J.}~\bibnamefont
  {Rueda}}, \bibinfo {author} {\bibfnamefont {R.}~\bibnamefont {Ruffini}},\
  and\ \bibinfo {author} {\bibfnamefont {R.}~\bibnamefont {Kerr}},\ }\href
  {https://doi.org/10.3847/1538-4357/ac5b6e} {\bibfield  {journal} {\bibinfo
  {journal} {Astrophys. J.}\ }\textbf {\bibinfo {volume} {929}},\ \bibinfo
  {pages} {56} (\bibinfo {year} {2022})}\BibitemShut {NoStop}%
\bibitem [{\citenamefont {Rastegarnia}\ \emph {et~al.}(2022)\citenamefont
  {Rastegarnia} \emph {et~al.}}]{rastegarnia2022structure}%
  \BibitemOpen
  \bibfield  {author} {\bibinfo {author} {\bibfnamefont {F.}~\bibnamefont
  {Rastegarnia}} \emph {et~al.},\ }\href
  {https://doi.org/10.1140/epjc/s10052-022-10750-x} {\bibfield  {journal}
  {\bibinfo  {journal} {Eur. Phys. J. C}\ }\textbf {\bibinfo {volume} {82}},\
  \bibinfo {pages} {778} (\bibinfo {year} {2022})}\BibitemShut {NoStop}%
\bibitem [{\citenamefont {Abramowski}\ \emph {et~al.}(2016)\citenamefont
  {Abramowski} \emph {et~al.}}]{collaboration2016acceleration}%
  \BibitemOpen
  \bibfield  {author} {\bibinfo {author} {\bibfnamefont {A.}~\bibnamefont
  {Abramowski}} \emph {et~al.} (\bibinfo {collaboration} {HESS
  Collaboration}),\ }\href {https://doi.org/10.1038/nature17147} {\bibfield
  {journal} {\bibinfo  {journal} {Nature (London)}\ }\textbf {\bibinfo {volume}
  {531}},\ \bibinfo {pages} {476} (\bibinfo {year} {2016})}\BibitemShut
  {NoStop}%
\bibitem [{\citenamefont {Aharonian}\ and\ \citenamefont
  {Sunyaev}(1984)}]{aharonian1984gamma}%
  \BibitemOpen
  \bibfield  {author} {\bibinfo {author} {\bibfnamefont {F.}~\bibnamefont
  {Aharonian}}\ and\ \bibinfo {author} {\bibfnamefont {R.}~\bibnamefont
  {Sunyaev}},\ }\href {https://doi.org/10.1093/mnras/210.2.257} {\bibfield
  {journal} {\bibinfo  {journal} {Mon. Not. R. Astron. Soc.}\ }\textbf
  {\bibinfo {volume} {210}},\ \bibinfo {pages} {257} (\bibinfo {year}
  {1984})}\BibitemShut {NoStop}%
\bibitem [{\citenamefont {Jean}\ and\ \citenamefont
  {Guessoum}(2001)}]{jean2001neutron}%
  \BibitemOpen
  \bibfield  {author} {\bibinfo {author} {\bibfnamefont {P.}~\bibnamefont
  {Jean}}\ and\ \bibinfo {author} {\bibfnamefont {N.}~\bibnamefont
  {Guessoum}},\ }\href {https://doi.org/10.1051/0004-6361:20011201} {\bibfield
  {journal} {\bibinfo  {journal} {Astron. Astrophys.}\ }\textbf {\bibinfo
  {volume} {378}},\ \bibinfo {pages} {509} (\bibinfo {year}
  {2001})}\BibitemShut {NoStop}%
\bibitem [{\citenamefont {Kafexhiu}\ \emph {et~al.}(2019)\citenamefont
  {Kafexhiu}, \citenamefont {Aharonian},\ and\ \citenamefont
  {Barkov}}]{kafexhiu2019nuclear}%
  \BibitemOpen
  \bibfield  {author} {\bibinfo {author} {\bibfnamefont {E.}~\bibnamefont
  {Kafexhiu}}, \bibinfo {author} {\bibfnamefont {F.}~\bibnamefont
  {Aharonian}},\ and\ \bibinfo {author} {\bibfnamefont {M.}~\bibnamefont
  {Barkov}},\ }\href {https://doi.org/10.1051/0004-6361/201833948} {\bibfield
  {journal} {\bibinfo  {journal} {Astron. Astrophys.}\ }\textbf {\bibinfo
  {volume} {623}},\ \bibinfo {pages} {A174} (\bibinfo {year}
  {2019})}\BibitemShut {NoStop}%
\bibitem [{\citenamefont {{Shapiro}}\ \emph {et~al.}(1976)\citenamefont
  {{Shapiro}}, \citenamefont {{Lightman}},\ and\ \citenamefont
  {{Eardley}}}]{1976ApJ...204..187S}%
  \BibitemOpen
  \bibfield  {author} {\bibinfo {author} {\bibfnamefont {S.~L.}\ \bibnamefont
  {{Shapiro}}}, \bibinfo {author} {\bibfnamefont {A.~P.}\ \bibnamefont
  {{Lightman}}},\ and\ \bibinfo {author} {\bibfnamefont {D.~M.}\ \bibnamefont
  {{Eardley}}},\ }\href {https://doi.org/10.1086/154162} {\bibfield  {journal}
  {\bibinfo  {journal} {Astrophys. J.}\ }\textbf {\bibinfo {volume} {204}},\
  \bibinfo {pages} {187} (\bibinfo {year} {1976})}\BibitemShut {NoStop}%
\bibitem [{\citenamefont {Park}(1995)}]{park1995stability}%
  \BibitemOpen
  \bibfield  {author} {\bibinfo {author} {\bibfnamefont {M.-G.}\ \bibnamefont
  {Park}},\ }\href {https://adsabs.harvard.edu/full/1995JKAS...28...97P}
  {\bibfield  {journal} {\bibinfo  {journal} {J. Korean Astron. Soc.}\ }\textbf
  {\bibinfo {volume} {28}},\ \bibinfo {pages} {97} (\bibinfo {year}
  {1995})}\BibitemShut {NoStop}%
\bibitem [{\citenamefont {Narayan}\ and\ \citenamefont
  {Yi}(1995{\natexlab{a}})}]{narayan1995advection}%
  \BibitemOpen
  \bibfield  {author} {\bibinfo {author} {\bibfnamefont {R.}~\bibnamefont
  {Narayan}}\ and\ \bibinfo {author} {\bibfnamefont {I.}~\bibnamefont {Yi}},\
  }\href {https://doi.org/10.1086/176343} {\bibfield  {journal} {\bibinfo
  {journal} {Astrophys. J.}\ }\textbf {\bibinfo {volume} {452}},\ \bibinfo
  {pages} {710} (\bibinfo {year} {1995}{\natexlab{a}})}\BibitemShut {NoStop}%
\bibitem [{\citenamefont {Narayan}\ and\ \citenamefont
  {Yi}(1994)}]{narayan1994advection}%
  \BibitemOpen
  \bibfield  {author} {\bibinfo {author} {\bibfnamefont {R.}~\bibnamefont
  {Narayan}}\ and\ \bibinfo {author} {\bibfnamefont {I.}~\bibnamefont {Yi}},\
  }\href {https://doi.org/10.1086/187381} {\bibfield  {journal} {\bibinfo
  {journal} {Astrophys. J.}\ }\textbf {\bibinfo {volume} {428}},\ \bibinfo
  {pages} {L13} (\bibinfo {year} {1994})}\BibitemShut {NoStop}%
\bibitem [{\citenamefont {Narayan}\ \emph
  {et~al.}(1997{\natexlab{a}})\citenamefont {Narayan}, \citenamefont {Barret},\
  and\ \citenamefont {McClintock}}]{narayan1997advection}%
  \BibitemOpen
  \bibfield  {author} {\bibinfo {author} {\bibfnamefont {R.}~\bibnamefont
  {Narayan}}, \bibinfo {author} {\bibfnamefont {D.}~\bibnamefont {Barret}},\
  and\ \bibinfo {author} {\bibfnamefont {J.~E.}\ \bibnamefont {McClintock}},\
  }\href {https://doi.org/10.1086/304134} {\bibfield  {journal} {\bibinfo
  {journal} {Astrophys. J.}\ }\textbf {\bibinfo {volume} {482}},\ \bibinfo
  {pages} {448} (\bibinfo {year} {1997}{\natexlab{a}})}\BibitemShut {NoStop}%
\bibitem [{\citenamefont {Narayan}\ \emph {et~al.}(1998)\citenamefont
  {Narayan}, \citenamefont {Mahadevan}, \citenamefont {Grindlay}, \citenamefont
  {Popham},\ and\ \citenamefont {Gammie}}]{narayan1998advection}%
  \BibitemOpen
  \bibfield  {author} {\bibinfo {author} {\bibfnamefont {R.}~\bibnamefont
  {Narayan}}, \bibinfo {author} {\bibfnamefont {R.}~\bibnamefont {Mahadevan}},
  \bibinfo {author} {\bibfnamefont {J.~E.}\ \bibnamefont {Grindlay}}, \bibinfo
  {author} {\bibfnamefont {R.~G.}\ \bibnamefont {Popham}},\ and\ \bibinfo
  {author} {\bibfnamefont {C.}~\bibnamefont {Gammie}},\ }\href
  {https://doi.org/10.1086/305070} {\bibfield  {journal} {\bibinfo  {journal}
  {Astrophys. J.}\ }\textbf {\bibinfo {volume} {492}},\ \bibinfo {pages} {554}
  (\bibinfo {year} {1998})}\BibitemShut {NoStop}%
\bibitem [{\citenamefont {Ptuskin}(2006)}]{ptuskin2006cosmic}%
  \BibitemOpen
  \bibfield  {author} {\bibinfo {author} {\bibfnamefont {V.}~\bibnamefont
  {Ptuskin}},\ }\href {https://doi.org/10.1088/1742-6596/47/1/014} {\bibfield
  {journal} {\bibinfo  {journal} {J. Phys. Conf. Ser.}\ }\textbf {\bibinfo
  {volume} {47}},\ \bibinfo {pages} {113} (\bibinfo {year} {2006})}\BibitemShut
  {NoStop}%
\bibitem [{\citenamefont {Paczyñsky}\ and\ \citenamefont
  {Wiita}(1980)}]{paczynsky1980thick}%
  \BibitemOpen
  \bibfield  {author} {\bibinfo {author} {\bibfnamefont {B.}~\bibnamefont
  {Paczyñsky}}\ and\ \bibinfo {author} {\bibfnamefont {P.~J.}\ \bibnamefont
  {Wiita}},\ }\href
  {https://ui.adsabs.harvard.edu/abs/1980A%26A....88...23P/abstract} {\bibfield
   {journal} {\bibinfo  {journal} {Astron. Astrophys.}\ }\textbf {\bibinfo
  {volume} {88}},\ \bibinfo {pages} {23} (\bibinfo {year} {1980})}\BibitemShut
  {NoStop}%
\bibitem [{\citenamefont {Narayan}\ \emph
  {et~al.}(1997{\natexlab{b}})\citenamefont {Narayan}, \citenamefont {Kato},\
  and\ \citenamefont {Honma}}]{narayan1997global}%
  \BibitemOpen
  \bibfield  {author} {\bibinfo {author} {\bibfnamefont {R.}~\bibnamefont
  {Narayan}}, \bibinfo {author} {\bibfnamefont {S.}~\bibnamefont {Kato}},\ and\
  \bibinfo {author} {\bibfnamefont {F.}~\bibnamefont {Honma}},\ }\href
  {https://doi.org/10.1086/303591} {\bibfield  {journal} {\bibinfo  {journal}
  {Astrophys. J.}\ }\textbf {\bibinfo {volume} {476}},\ \bibinfo {pages} {49}
  (\bibinfo {year} {1997}{\natexlab{b}})}\BibitemShut {NoStop}%
\bibitem [{\citenamefont {Gammie}\ and\ \citenamefont
  {Popham}(1998)}]{gammie1998advection}%
  \BibitemOpen
  \bibfield  {author} {\bibinfo {author} {\bibfnamefont {C.~F.}\ \bibnamefont
  {Gammie}}\ and\ \bibinfo {author} {\bibfnamefont {R.}~\bibnamefont
  {Popham}},\ }\href {https://doi.org/10.1086/305521} {\bibfield  {journal}
  {\bibinfo  {journal} {Astrophys. J.}\ }\textbf {\bibinfo {volume} {498}},\
  \bibinfo {pages} {313} (\bibinfo {year} {1998})}\BibitemShut {NoStop}%
\bibitem [{\citenamefont {Popham}\ and\ \citenamefont
  {Gammie}(1998)}]{popham1998advection}%
  \BibitemOpen
  \bibfield  {author} {\bibinfo {author} {\bibfnamefont {R.}~\bibnamefont
  {Popham}}\ and\ \bibinfo {author} {\bibfnamefont {C.~F.}\ \bibnamefont
  {Gammie}},\ }\href {https://doi.org/10.1086/306054} {\bibfield  {journal}
  {\bibinfo  {journal} {Astrophys. J.}\ }\textbf {\bibinfo {volume} {504}},\
  \bibinfo {pages} {419} (\bibinfo {year} {1998})}\BibitemShut {NoStop}%
\bibitem [{\citenamefont {Agol}(2000)}]{agol2000sagittarius}%
  \BibitemOpen
  \bibfield  {author} {\bibinfo {author} {\bibfnamefont {E.}~\bibnamefont
  {Agol}},\ }\href {https://doi.org/10.1086/31281} {\bibfield  {journal}
  {\bibinfo  {journal} {Astrophys. J.}\ }\textbf {\bibinfo {volume} {538}},\
  \bibinfo {pages} {L121} (\bibinfo {year} {2000})}\BibitemShut {NoStop}%
\bibitem [{\citenamefont {Marrone}\ \emph {et~al.}(2006)\citenamefont
  {Marrone}, \citenamefont {Moran}, \citenamefont {Zhao},\ and\ \citenamefont
  {Rao}}]{marrone2006unambiguous}%
  \BibitemOpen
  \bibfield  {author} {\bibinfo {author} {\bibfnamefont {D.~P.}\ \bibnamefont
  {Marrone}}, \bibinfo {author} {\bibfnamefont {J.~M.}\ \bibnamefont {Moran}},
  \bibinfo {author} {\bibfnamefont {J.-H.}\ \bibnamefont {Zhao}},\ and\
  \bibinfo {author} {\bibfnamefont {R.}~\bibnamefont {Rao}},\ }\href
  {https://doi.org/10.1086/510850} {\bibfield  {journal} {\bibinfo  {journal}
  {Astrophys. J.}\ }\textbf {\bibinfo {volume} {654}},\ \bibinfo {pages} {L57}
  (\bibinfo {year} {2006})}\BibitemShut {NoStop}%
\bibitem [{\citenamefont {Sharma}\ \emph {et~al.}(2007)\citenamefont {Sharma},
  \citenamefont {Quataert},\ and\ \citenamefont {Stone}}]{sharma2007faraday}%
  \BibitemOpen
  \bibfield  {author} {\bibinfo {author} {\bibfnamefont {P.}~\bibnamefont
  {Sharma}}, \bibinfo {author} {\bibfnamefont {E.}~\bibnamefont {Quataert}},\
  and\ \bibinfo {author} {\bibfnamefont {J.~M.}\ \bibnamefont {Stone}},\ }\href
  {https://doi.org/10.1086/523267} {\bibfield  {journal} {\bibinfo  {journal}
  {Astrophys. J.}\ }\textbf {\bibinfo {volume} {671}},\ \bibinfo {pages} {1696}
  (\bibinfo {year} {2007})}\BibitemShut {NoStop}%
\bibitem [{\citenamefont {Shcherbakov}\ \emph {et~al.}(2012)\citenamefont
  {Shcherbakov}, \citenamefont {Penna},\ and\ \citenamefont
  {McKinney}}]{shcherbakov2012sagittarius}%
  \BibitemOpen
  \bibfield  {author} {\bibinfo {author} {\bibfnamefont {R.~V.}\ \bibnamefont
  {Shcherbakov}}, \bibinfo {author} {\bibfnamefont {R.~F.}\ \bibnamefont
  {Penna}},\ and\ \bibinfo {author} {\bibfnamefont {J.~C.}\ \bibnamefont
  {McKinney}},\ }\href {https://doi.org/10.1088/0004-637X/755/2/133} {\bibfield
   {journal} {\bibinfo  {journal} {Astrophys. J.}\ }\textbf {\bibinfo {volume}
  {755}},\ \bibinfo {pages} {133} (\bibinfo {year} {2012})}\BibitemShut
  {NoStop}%
\bibitem [{\citenamefont {Bower}\ \emph {et~al.}(2018)\citenamefont {Bower}
  \emph {et~al.}}]{bower2018alma}%
  \BibitemOpen
  \bibfield  {author} {\bibinfo {author} {\bibfnamefont {G.~C.}\ \bibnamefont
  {Bower}} \emph {et~al.},\ }\href {https://doi.org/10.3847/1538-4357/aae983}
  {\bibfield  {journal} {\bibinfo  {journal} {Astrophys. J.}\ }\textbf
  {\bibinfo {volume} {868}},\ \bibinfo {pages} {101} (\bibinfo {year}
  {2018})}\BibitemShut {NoStop}%
\bibitem [{\citenamefont {{\"O}zel}\ \emph {et~al.}(2000)\citenamefont
  {{\"O}zel}, \citenamefont {Psaltis},\ and\ \citenamefont
  {Narayan}}]{ozel2000hybrid}%
  \BibitemOpen
  \bibfield  {author} {\bibinfo {author} {\bibfnamefont {F.}~\bibnamefont
  {{\"O}zel}}, \bibinfo {author} {\bibfnamefont {D.}~\bibnamefont {Psaltis}},\
  and\ \bibinfo {author} {\bibfnamefont {R.}~\bibnamefont {Narayan}},\ }\href
  {https://doi.org/10.1086/309396} {\bibfield  {journal} {\bibinfo  {journal}
  {Astrophys. J.}\ }\textbf {\bibinfo {volume} {541}},\ \bibinfo {pages} {234}
  (\bibinfo {year} {2000})}\BibitemShut {NoStop}%
\bibitem [{\citenamefont {Weaver}(1976)}]{weaver1976reaction}%
  \BibitemOpen
  \bibfield  {author} {\bibinfo {author} {\bibfnamefont {T.~A.}\ \bibnamefont
  {Weaver}},\ }\href {https://doi.org/10.1103/PhysRevA.13.1563} {\bibfield
  {journal} {\bibinfo  {journal} {Phys. Rev. A}\ }\textbf {\bibinfo {volume}
  {13}},\ \bibinfo {pages} {1563} (\bibinfo {year} {1976})}\BibitemShut
  {NoStop}%
\bibitem [{\citenamefont {Meyer}(1972)}]{meyer1972deuterons}%
  \BibitemOpen
  \bibfield  {author} {\bibinfo {author} {\bibfnamefont {J.}~\bibnamefont
  {Meyer}},\ }\href
  {https://ui.adsabs.harvard.edu/abs/1972A%26AS....7..417M/abstract} {\bibfield
   {journal} {\bibinfo  {journal} {Astron. Astrophys. Suppl. Ser.}\ }\textbf
  {\bibinfo {volume} {7}},\ \bibinfo {pages} {417} (\bibinfo {year}
  {1972})}\BibitemShut {NoStop}%
\bibitem [{\citenamefont {Jung}\ \emph {et~al.}(1973)\citenamefont {Jung} \emph
  {et~al.}}]{jung1973proton}%
  \BibitemOpen
  \bibfield  {author} {\bibinfo {author} {\bibfnamefont {M.}~\bibnamefont
  {Jung}} \emph {et~al.},\ }\href {https://doi.org/10.1103/PhysRevC.7.2209}
  {\bibfield  {journal} {\bibinfo  {journal} {Phys. Rev. C}\ }\textbf {\bibinfo
  {volume} {7}},\ \bibinfo {pages} {2209} (\bibinfo {year} {1973})}\BibitemShut
  {NoStop}%
\bibitem [{\citenamefont {King}\ \emph {et~al.}(1977)\citenamefont {King},
  \citenamefont {Austin}, \citenamefont {Rossner},\ and\ \citenamefont
  {Chien}}]{king1977li}%
  \BibitemOpen
  \bibfield  {author} {\bibinfo {author} {\bibfnamefont {C.}~\bibnamefont
  {King}}, \bibinfo {author} {\bibfnamefont {S.~M.}\ \bibnamefont {Austin}},
  \bibinfo {author} {\bibfnamefont {H.}~\bibnamefont {Rossner}},\ and\ \bibinfo
  {author} {\bibfnamefont {W.}~\bibnamefont {Chien}},\ }\href
  {https://doi.org/10.1103/PhysRevC.16.1712} {\bibfield  {journal} {\bibinfo
  {journal} {Phys. Rev. C}\ }\textbf {\bibinfo {volume} {16}},\ \bibinfo
  {pages} {1712} (\bibinfo {year} {1977})}\BibitemShut {NoStop}%
\bibitem [{\citenamefont {Yiou}\ \emph {et~al.}(1977)\citenamefont {Yiou},
  \citenamefont {Raisbeck},\ and\ \citenamefont {Quechon}}]{yiou1977cross}%
  \BibitemOpen
  \bibfield  {author} {\bibinfo {author} {\bibfnamefont {F.}~\bibnamefont
  {Yiou}}, \bibinfo {author} {\bibfnamefont {G.}~\bibnamefont {Raisbeck}},\
  and\ \bibinfo {author} {\bibfnamefont {H.}~\bibnamefont {Quechon}},\ }in\
  \href {https://ui.adsabs.harvard.edu/abs/1977ICRC....2...72Y/abstract} {\emph
  {\bibinfo {booktitle} {International Cosmic Ray Conference}}},\ Vol.~\bibinfo
  {volume} {2}\ (\bibinfo {organization} {Hungarian Academy of Sciences,
  Budapest},\ \bibinfo {year} {1977})\ p.~\bibinfo {pages} {72}\BibitemShut
  {NoStop}%
\bibitem [{\citenamefont {Woo}\ \emph {et~al.}(1985)\citenamefont {Woo},
  \citenamefont {Kwiatkowski}, \citenamefont {Zhou},\ and\ \citenamefont
  {Viola}}]{woo1985cross}%
  \BibitemOpen
  \bibfield  {author} {\bibinfo {author} {\bibfnamefont {L.}~\bibnamefont
  {Woo}}, \bibinfo {author} {\bibfnamefont {K.}~\bibnamefont {Kwiatkowski}},
  \bibinfo {author} {\bibfnamefont {S.}~\bibnamefont {Zhou}},\ and\ \bibinfo
  {author} {\bibfnamefont {V.}~\bibnamefont {Viola}},\ }\href
  {https://doi.org/10.1103/PhysRevC.32.706} {\bibfield  {journal} {\bibinfo
  {journal} {Phys. Rev. C}\ }\textbf {\bibinfo {volume} {32}},\ \bibinfo
  {pages} {706} (\bibinfo {year} {1985})}\BibitemShut {NoStop}%
\bibitem [{\citenamefont {Mercer}\ \emph {et~al.}(1997)\citenamefont {Mercer},
  \citenamefont {Austin},\ and\ \citenamefont {Glagola}}]{mercer1997suggested}%
  \BibitemOpen
  \bibfield  {author} {\bibinfo {author} {\bibfnamefont {D.}~\bibnamefont
  {Mercer}}, \bibinfo {author} {\bibfnamefont {S.~M.}\ \bibnamefont {Austin}},\
  and\ \bibinfo {author} {\bibfnamefont {B.}~\bibnamefont {Glagola}},\ }\href
  {https://doi.org/10.1103/PhysRevC.55.946} {\bibfield  {journal} {\bibinfo
  {journal} {Phys. Rev. C}\ }\textbf {\bibinfo {volume} {55}},\ \bibinfo
  {pages} {946} (\bibinfo {year} {1997})}\BibitemShut {NoStop}%
\bibitem [{\citenamefont {Mercer}\ \emph {et~al.}(2001)\citenamefont {Mercer}
  \emph {et~al.}}]{mercer2001production}%
  \BibitemOpen
  \bibfield  {author} {\bibinfo {author} {\bibfnamefont {D.~J.}\ \bibnamefont
  {Mercer}} \emph {et~al.},\ }\href
  {https://doi.org/10.1103/PhysRevC.63.065805} {\bibfield  {journal} {\bibinfo
  {journal} {Phys. Rev. C}\ }\textbf {\bibinfo {volume} {63}},\ \bibinfo
  {pages} {065805} (\bibinfo {year} {2001})}\BibitemShut {NoStop}%
\bibitem [{\citenamefont {Kawabata}\ \emph {et~al.}(2017)\citenamefont
  {Kawabata} \emph {et~al.}}]{kawabata2017time}%
  \BibitemOpen
  \bibfield  {author} {\bibinfo {author} {\bibfnamefont {T.}~\bibnamefont
  {Kawabata}} \emph {et~al.},\ }\href
  {https://doi.org/10.1103/PhysRevLett.118.052701} {\bibfield  {journal}
  {\bibinfo  {journal} {Phys. Rev. Lett.}\ }\textbf {\bibinfo {volume} {118}},\
  \bibinfo {pages} {052701} (\bibinfo {year} {2017})}\BibitemShut {NoStop}%
\bibitem [{\citenamefont {Kelner}\ \emph {et~al.}(2006)\citenamefont {Kelner},
  \citenamefont {Aharonian},\ and\ \citenamefont {Bugayov}}]{kelner2006energy}%
  \BibitemOpen
  \bibfield  {author} {\bibinfo {author} {\bibfnamefont {S.~R.}\ \bibnamefont
  {Kelner}}, \bibinfo {author} {\bibfnamefont {F.~A.}\ \bibnamefont
  {Aharonian}},\ and\ \bibinfo {author} {\bibfnamefont {V.~V.}\ \bibnamefont
  {Bugayov}},\ }\href {https://doi.org/10.1103/PhysRevD.74.034018} {\bibfield
  {journal} {\bibinfo  {journal} {Phys. Rev. D}\ }\textbf {\bibinfo {volume}
  {74}},\ \bibinfo {pages} {034018} (\bibinfo {year} {2006})}\BibitemShut
  {NoStop}%
\bibitem [{\citenamefont {Guessoum}\ and\ \citenamefont
  {Kazanas}(1990)}]{guessoum1990neutron}%
  \BibitemOpen
  \bibfield  {author} {\bibinfo {author} {\bibfnamefont {N.}~\bibnamefont
  {Guessoum}}\ and\ \bibinfo {author} {\bibfnamefont {D.}~\bibnamefont
  {Kazanas}},\ }\href {https://doi.org/10.1086/169005} {\bibfield  {journal}
  {\bibinfo  {journal} {Astrophys. J.}\ }\textbf {\bibinfo {volume} {358}},\
  \bibinfo {pages} {525} (\bibinfo {year} {1990})}\BibitemShut {NoStop}%
\bibitem [{\citenamefont {Brysk}(1973)}]{brysk1973fusion}%
  \BibitemOpen
  \bibfield  {author} {\bibinfo {author} {\bibfnamefont {H.}~\bibnamefont
  {Brysk}},\ }\href {https://doi.org/10.1088/0032-1028/15/7/001} {\bibfield
  {journal} {\bibinfo  {journal} {Plasma Phys.}\ }\textbf {\bibinfo {volume}
  {15}},\ \bibinfo {pages} {611} (\bibinfo {year} {1973})}\BibitemShut
  {NoStop}%
\bibitem [{\citenamefont {Zyla}\ \emph {et~al.}(2020)\citenamefont {Zyla} \emph
  {et~al.}}]{particle2020review}%
  \BibitemOpen
  \bibfield  {author} {\bibinfo {author} {\bibfnamefont {P.}~\bibnamefont
  {Zyla}} \emph {et~al.} (\bibinfo {collaboration} {Particle Data Group}),\
  }\href {https://doi.org/10.1093/ptep/ptaa104} {\bibfield  {journal} {\bibinfo
   {journal} {Prog. Theor. Exp. Phys.}\ }\textbf {\bibinfo {volume} {2020}},\
  \bibinfo {pages} {083C01} (\bibinfo {year} {2020})}\BibitemShut {NoStop}%
\bibitem [{\citenamefont {Goldreich}\ and\ \citenamefont
  {Julian}(1969)}]{goldreich1969pulsar}%
  \BibitemOpen
  \bibfield  {author} {\bibinfo {author} {\bibfnamefont {P.}~\bibnamefont
  {Goldreich}}\ and\ \bibinfo {author} {\bibfnamefont {W.~H.}\ \bibnamefont
  {Julian}},\ }\href {https://doi.org/10.1086/150119} {\bibfield  {journal}
  {\bibinfo  {journal} {Astrophys. J.}\ }\textbf {\bibinfo {volume} {157}},\
  \bibinfo {pages} {869} (\bibinfo {year} {1969})}\BibitemShut {NoStop}%
\bibitem [{\citenamefont {Akiyama}\ \emph {et~al.}(2022)\citenamefont {Akiyama}
  \emph {et~al.}}]{akiyama2022first}%
  \BibitemOpen
  \bibfield  {author} {\bibinfo {author} {\bibfnamefont {K.}~\bibnamefont
  {Akiyama}} \emph {et~al.} (\bibinfo {collaboration} {Event Horizon Telescope
  Collaboration}),\ }\href {https://doi.org/10.3847/2041-8213/ac6672}
  {\bibfield  {journal} {\bibinfo  {journal} {Astrophys. J. Lett.}\ }\textbf
  {\bibinfo {volume} {930}},\ \bibinfo {pages} {L16} (\bibinfo {year}
  {2022})}\BibitemShut {NoStop}%
\bibitem [{\citenamefont {Narayan}\ and\ \citenamefont
  {Yi}(1995{\natexlab{b}})}]{narayan1995advection2}%
  \BibitemOpen
  \bibfield  {author} {\bibinfo {author} {\bibfnamefont {R.}~\bibnamefont
  {Narayan}}\ and\ \bibinfo {author} {\bibfnamefont {I.}~\bibnamefont {Yi}},\
  }\href {https://doi.org/10.1086/175599} {\bibfield  {journal} {\bibinfo
  {journal} {Astrophys. J.}\ }\textbf {\bibinfo {volume} {444}},\ \bibinfo
  {pages} {231} (\bibinfo {year} {1995}{\natexlab{b}})}\BibitemShut {NoStop}%
\bibitem [{\citenamefont {Park}\ and\ \citenamefont
  {Ostriker}(1999)}]{park1999thermal}%
  \BibitemOpen
  \bibfield  {author} {\bibinfo {author} {\bibfnamefont {M.-G.}\ \bibnamefont
  {Park}}\ and\ \bibinfo {author} {\bibfnamefont {J.~P.}\ \bibnamefont
  {Ostriker}},\ }\href {https://doi.org/10.1086/308061} {\bibfield  {journal}
  {\bibinfo  {journal} {Astrophys. J.}\ }\textbf {\bibinfo {volume} {527}},\
  \bibinfo {pages} {247} (\bibinfo {year} {1999})}\BibitemShut {NoStop}%
\bibitem [{\citenamefont {Park}\ and\ \citenamefont
  {Ostriker}(2007)}]{park2007compton}%
  \BibitemOpen
  \bibfield  {author} {\bibinfo {author} {\bibfnamefont {M.-G.}\ \bibnamefont
  {Park}}\ and\ \bibinfo {author} {\bibfnamefont {J.~P.}\ \bibnamefont
  {Ostriker}},\ }\href {https://doi.org/10.1086/509698} {\bibfield  {journal}
  {\bibinfo  {journal} {Astrophys. J.}\ }\textbf {\bibinfo {volume} {655}},\
  \bibinfo {pages} {88} (\bibinfo {year} {2007})}\BibitemShut {NoStop}%
\bibitem [{\citenamefont {Mo{\'s}cibrodzka}\ \emph {et~al.}(2011)\citenamefont
  {Mo{\'s}cibrodzka}, \citenamefont {Gammie}, \citenamefont {Dolence},\ and\
  \citenamefont {Shiokawa}}]{moscibrodzka2011pair}%
  \BibitemOpen
  \bibfield  {author} {\bibinfo {author} {\bibfnamefont {M.}~\bibnamefont
  {Mo{\'s}cibrodzka}}, \bibinfo {author} {\bibfnamefont {C.~F.}\ \bibnamefont
  {Gammie}}, \bibinfo {author} {\bibfnamefont {J.~C.}\ \bibnamefont
  {Dolence}},\ and\ \bibinfo {author} {\bibfnamefont {H.}~\bibnamefont
  {Shiokawa}},\ }\href {https://doi.org/10.1088/0004-637X/735/1/9} {\bibfield
  {journal} {\bibinfo  {journal} {Astrophys. J.}\ }\textbf {\bibinfo {volume}
  {735}},\ \bibinfo {pages} {9} (\bibinfo {year} {2011})}\BibitemShut {NoStop}%
\bibitem [{\citenamefont {Wong}\ \emph {et~al.}(2021)\citenamefont {Wong},
  \citenamefont {Ryan},\ and\ \citenamefont {Gammie}}]{wong2021pair}%
  \BibitemOpen
  \bibfield  {author} {\bibinfo {author} {\bibfnamefont {G.~N.}\ \bibnamefont
  {Wong}}, \bibinfo {author} {\bibfnamefont {B.~R.}\ \bibnamefont {Ryan}},\
  and\ \bibinfo {author} {\bibfnamefont {C.~F.}\ \bibnamefont {Gammie}},\
  }\href {https://doi.org/10.3847/1538-4357/abd0f9} {\bibfield  {journal}
  {\bibinfo  {journal} {Astrophys. J.}\ }\textbf {\bibinfo {volume} {907}},\
  \bibinfo {pages} {73} (\bibinfo {year} {2021})}\BibitemShut {NoStop}%
\bibitem [{\citenamefont {Zaja{\v{c}}ek}\ \emph {et~al.}(2018)\citenamefont
  {Zaja{\v{c}}ek}, \citenamefont {Tursunov}, \citenamefont {Eckart},\ and\
  \citenamefont {Britzen}}]{zajavcek2018charge}%
  \BibitemOpen
  \bibfield  {author} {\bibinfo {author} {\bibfnamefont {M.}~\bibnamefont
  {Zaja{\v{c}}ek}}, \bibinfo {author} {\bibfnamefont {A.}~\bibnamefont
  {Tursunov}}, \bibinfo {author} {\bibfnamefont {A.}~\bibnamefont {Eckart}},\
  and\ \bibinfo {author} {\bibfnamefont {S.}~\bibnamefont {Britzen}},\ }\href
  {https://doi.org/10.1093/mnras/sty2182} {\bibfield  {journal} {\bibinfo
  {journal} {Mon. Not. R. Astron. Soc.}\ }\textbf {\bibinfo {volume} {480}},\
  \bibinfo {pages} {4408} (\bibinfo {year} {2018})}\BibitemShut {NoStop}%
\bibitem [{\citenamefont {Zajacek}\ and\ \citenamefont
  {Tursunov}(2019)}]{zajavcek2019electric}%
  \BibitemOpen
  \bibfield  {author} {\bibinfo {author} {\bibfnamefont {M.}~\bibnamefont
  {Zajacek}}\ and\ \bibinfo {author} {\bibfnamefont {A.}~\bibnamefont
  {Tursunov}},\ }\href
  {https://ui.adsabs.harvard.edu/abs/2019Obs...139..231Z/abstract} {\bibfield
  {journal} {\bibinfo  {journal} {Observatory}\ }\textbf {\bibinfo {volume}
  {139}},\ \bibinfo {pages} {231} (\bibinfo {year} {2019})}\BibitemShut
  {NoStop}%
\bibitem [{\citenamefont {Eckart}\ \emph {et~al.}(2017)\citenamefont {Eckart}
  \emph {et~al.}}]{eckart2017milky}%
  \BibitemOpen
  \bibfield  {author} {\bibinfo {author} {\bibfnamefont {A.}~\bibnamefont
  {Eckart}} \emph {et~al.},\ }\href {https://doi.org/10.1007/s10701-017-0079-2}
  {\bibfield  {journal} {\bibinfo  {journal} {Found. Phys.}\ }\textbf {\bibinfo
  {volume} {47}},\ \bibinfo {pages} {553} (\bibinfo {year} {2017})}\BibitemShut
  {NoStop}%
\bibitem [{\citenamefont {Eatough}\ \emph {et~al.}(2013)\citenamefont {Eatough}
  \emph {et~al.}}]{eatough2013strong}%
  \BibitemOpen
  \bibfield  {author} {\bibinfo {author} {\bibfnamefont {R.}~\bibnamefont
  {Eatough}} \emph {et~al.},\ }\href {https://doi.org/10.1038/nature12499}
  {\bibfield  {journal} {\bibinfo  {journal} {Nature (London)}\ }\textbf
  {\bibinfo {volume} {501}},\ \bibinfo {pages} {391} (\bibinfo {year}
  {2013})}\BibitemShut {NoStop}%
\bibitem [{\citenamefont {Cho}\ \emph {et~al.}(2022)\citenamefont {Cho} \emph
  {et~al.}}]{cho2022intrinsic}%
  \BibitemOpen
  \bibfield  {author} {\bibinfo {author} {\bibfnamefont {I.}~\bibnamefont
  {Cho}} \emph {et~al.},\ }\href {https://doi.org/10.3847/1538-4357/ac4165}
  {\bibfield  {journal} {\bibinfo  {journal} {Astrophys. J.}\ }\textbf
  {\bibinfo {volume} {926}},\ \bibinfo {pages} {108} (\bibinfo {year}
  {2022})}\BibitemShut {NoStop}%
\bibitem [{\citenamefont {Stuchl{\'\i}k}\ \emph {et~al.}(2021)\citenamefont
  {Stuchl{\'\i}k}, \citenamefont {Kolo{\v{s}}},\ and\ \citenamefont
  {Tursunov}}]{stuchlik2021penrose}%
  \BibitemOpen
  \bibfield  {author} {\bibinfo {author} {\bibfnamefont {Z.}~\bibnamefont
  {Stuchl{\'\i}k}}, \bibinfo {author} {\bibfnamefont {M.}~\bibnamefont
  {Kolo{\v{s}}}},\ and\ \bibinfo {author} {\bibfnamefont {A.}~\bibnamefont
  {Tursunov}},\ }\href {https://doi.org/10.3390/universe7110416} {\bibfield
  {journal} {\bibinfo  {journal} {Universe}\ }\textbf {\bibinfo {volume} {7}},\
  \bibinfo {pages} {416} (\bibinfo {year} {2021})}\BibitemShut {NoStop}%
\bibitem [{\citenamefont {Rybicki}\ and\ \citenamefont
  {Lightman}(1991)}]{rybicki1991radiative}%
  \BibitemOpen
  \bibfield  {author} {\bibinfo {author} {\bibfnamefont {G.~B.}\ \bibnamefont
  {Rybicki}}\ and\ \bibinfo {author} {\bibfnamefont {A.~P.}\ \bibnamefont
  {Lightman}},\ }\href@noop {} {\emph {\bibinfo {title} {Radiative Processes in
  Astrophysics}}}\ (\bibinfo  {publisher} {John Wiley \& Sons},\ \bibinfo
  {address} {New York},\ \bibinfo {year} {1991})\BibitemShut {NoStop}%
\bibitem [{\citenamefont {Ponti}\ \emph {et~al.}(2017)\citenamefont {Ponti}
  \emph {et~al.}}]{ponti2017powerful}%
  \BibitemOpen
  \bibfield  {author} {\bibinfo {author} {\bibfnamefont {G.}~\bibnamefont
  {Ponti}} \emph {et~al.},\ }\href {https://doi.org/10.1093/mnras/stx596}
  {\bibfield  {journal} {\bibinfo  {journal} {Mon. Not. R. Astron. Soc.}\
  }\textbf {\bibinfo {volume} {468}},\ \bibinfo {pages} {2447} (\bibinfo {year}
  {2017})}\BibitemShut {NoStop}%
\bibitem [{\citenamefont {Yuan}\ \emph {et~al.}(2004)\citenamefont {Yuan},
  \citenamefont {Quataert},\ and\ \citenamefont {Narayan}}]{yuan2004nature}%
  \BibitemOpen
  \bibfield  {author} {\bibinfo {author} {\bibfnamefont {F.}~\bibnamefont
  {Yuan}}, \bibinfo {author} {\bibfnamefont {E.}~\bibnamefont {Quataert}},\
  and\ \bibinfo {author} {\bibfnamefont {R.}~\bibnamefont {Narayan}},\ }\href
  {https://doi.org/10.1086/383117} {\bibfield  {journal} {\bibinfo  {journal}
  {Astrophys. J.}\ }\textbf {\bibinfo {volume} {606}},\ \bibinfo {pages} {894}
  (\bibinfo {year} {2004})}\BibitemShut {NoStop}%
\bibitem [{\citenamefont {Baring}\ \emph {et~al.}(2000)\citenamefont {Baring},
  \citenamefont {Jones},\ and\ \citenamefont {Ellison}}]{baring2000inverse}%
  \BibitemOpen
  \bibfield  {author} {\bibinfo {author} {\bibfnamefont {M.~G.}\ \bibnamefont
  {Baring}}, \bibinfo {author} {\bibfnamefont {F.~C.}\ \bibnamefont {Jones}},\
  and\ \bibinfo {author} {\bibfnamefont {D.~C.}\ \bibnamefont {Ellison}},\
  }\href {https://doi.org/10.1086/308194} {\bibfield  {journal} {\bibinfo
  {journal} {Astrophys. J.}\ }\textbf {\bibinfo {volume} {528}},\ \bibinfo
  {pages} {776} (\bibinfo {year} {2000})}\BibitemShut {NoStop}%
\bibitem [{\citenamefont {Tsuchiya}\ \emph {et~al.}(2004)\citenamefont
  {Tsuchiya} \emph {et~al.}}]{tsuchiya2004detection}%
  \BibitemOpen
  \bibfield  {author} {\bibinfo {author} {\bibfnamefont {K.}~\bibnamefont
  {Tsuchiya}} \emph {et~al.},\ }\href {https://doi.org/10.1086/421292}
  {\bibfield  {journal} {\bibinfo  {journal} {Astrophys. J.}\ }\textbf
  {\bibinfo {volume} {606}},\ \bibinfo {pages} {L115} (\bibinfo {year}
  {2004})}\BibitemShut {NoStop}%
\bibitem [{\citenamefont {Kosack}\ \emph {et~al.}(2004)\citenamefont {Kosack}
  \emph {et~al.}}]{kosack2004tev}%
  \BibitemOpen
  \bibfield  {author} {\bibinfo {author} {\bibfnamefont {K.}~\bibnamefont
  {Kosack}} \emph {et~al.},\ }\href {https://doi.org/10.1086/422469} {\bibfield
   {journal} {\bibinfo  {journal} {Astrophys. J.}\ }\textbf {\bibinfo {volume}
  {608}},\ \bibinfo {pages} {L97} (\bibinfo {year} {2004})}\BibitemShut
  {NoStop}%
\bibitem [{\citenamefont {Aharonian}\ \emph {et~al.}(2004)\citenamefont
  {Aharonian} \emph {et~al.}}]{aharonian2004very}%
  \BibitemOpen
  \bibfield  {author} {\bibinfo {author} {\bibfnamefont {F.}~\bibnamefont
  {Aharonian}} \emph {et~al.},\ }\href
  {https://doi.org/10.1051/0004-6361:200400055} {\bibfield  {journal} {\bibinfo
   {journal} {Astron. Astrophys.}\ }\textbf {\bibinfo {volume} {425}},\
  \bibinfo {pages} {L13} (\bibinfo {year} {2004})}\BibitemShut {NoStop}%
\bibitem [{\citenamefont {Albert}\ \emph {et~al.}(2006)\citenamefont {Albert}
  \emph {et~al.}}]{albert2006observation}%
  \BibitemOpen
  \bibfield  {author} {\bibinfo {author} {\bibfnamefont {J.}~\bibnamefont
  {Albert}} \emph {et~al.},\ }\href {https://doi.org/10.1086/501164} {\bibfield
   {journal} {\bibinfo  {journal} {Astrophys. J.}\ }\textbf {\bibinfo {volume}
  {638}},\ \bibinfo {pages} {L101} (\bibinfo {year} {2006})}\BibitemShut
  {NoStop}%
\bibitem [{\citenamefont {Aharonian}\ \emph {et~al.}(2009)\citenamefont
  {Aharonian} \emph {et~al.}}]{aharonian2009spectrum}%
  \BibitemOpen
  \bibfield  {author} {\bibinfo {author} {\bibfnamefont {F.}~\bibnamefont
  {Aharonian}} \emph {et~al.},\ }\href
  {https://doi.org/10.1051/0004-6361/200811569} {\bibfield  {journal} {\bibinfo
   {journal} {Astron. Astrophys.}\ }\textbf {\bibinfo {volume} {503}},\
  \bibinfo {pages} {817} (\bibinfo {year} {2009})}\BibitemShut {NoStop}%
\bibitem [{\citenamefont {Abdalla}\ \emph {et~al.}(2018)\citenamefont {Abdalla}
  \emph {et~al.}}]{abdalla2018characterising}%
  \BibitemOpen
  \bibfield  {author} {\bibinfo {author} {\bibfnamefont {H.}~\bibnamefont
  {Abdalla}} \emph {et~al.} (\bibinfo {collaboration} {HESS Collaboration}),\
  }\href {https://doi.org/10.1051/0004-6361/201730824} {\bibfield  {journal}
  {\bibinfo  {journal} {Astron. Astrophys.}\ }\textbf {\bibinfo {volume}
  {612}},\ \bibinfo {pages} {A9} (\bibinfo {year} {2018})}\BibitemShut
  {NoStop}%
\bibitem [{\citenamefont {Acciari}\ \emph {et~al.}(2020)\citenamefont {Acciari}
  \emph {et~al.}}]{acciari2020magic}%
  \BibitemOpen
  \bibfield  {author} {\bibinfo {author} {\bibfnamefont {V.}~\bibnamefont
  {Acciari}} \emph {et~al.} (\bibinfo {collaboration} {MAGIC Collaboration}),\
  }\href {https://doi.org/10.1051/0004-6361/201936896} {\bibfield  {journal}
  {\bibinfo  {journal} {Astron. Astrophys.}\ }\textbf {\bibinfo {volume}
  {642}},\ \bibinfo {pages} {A190} (\bibinfo {year} {2020})}\BibitemShut
  {NoStop}%
\bibitem [{\citenamefont {Adams}\ \emph {et~al.}(2021)\citenamefont {Adams}
  \emph {et~al.}}]{adams2021veritas}%
  \BibitemOpen
  \bibfield  {author} {\bibinfo {author} {\bibfnamefont {C.}~\bibnamefont
  {Adams}} \emph {et~al.},\ }\href {https://doi.org/10.3847/1538-4357/abf926}
  {\bibfield  {journal} {\bibinfo  {journal} {Astrophys. J.}\ }\textbf
  {\bibinfo {volume} {913}},\ \bibinfo {pages} {115} (\bibinfo {year}
  {2021})}\BibitemShut {NoStop}%
\bibitem [{\citenamefont {Dahmen}\ \emph {et~al.}(1998)\citenamefont {Dahmen},
  \citenamefont {Huttemeister}, \citenamefont {Wilson},\ and\ \citenamefont
  {Mauersberger}}]{dahmen1998molecular}%
  \BibitemOpen
  \bibfield  {author} {\bibinfo {author} {\bibfnamefont {G.}~\bibnamefont
  {Dahmen}}, \bibinfo {author} {\bibfnamefont {S.}~\bibnamefont
  {Huttemeister}}, \bibinfo {author} {\bibfnamefont {T.}~\bibnamefont
  {Wilson}},\ and\ \bibinfo {author} {\bibfnamefont {R.}~\bibnamefont
  {Mauersberger}},\ }\href
  {https://ui.adsabs.harvard.edu/abs/1998A%26A...331..959D/abstract} {\bibfield
   {journal} {\bibinfo  {journal} {Astron. Astrophys.}\ }\textbf {\bibinfo
  {volume} {331}},\ \bibinfo {pages} {959} (\bibinfo {year}
  {1998})}\BibitemShut {NoStop}%
\bibitem [{\citenamefont {Ferriere}\ \emph {et~al.}(2007)\citenamefont
  {Ferriere}, \citenamefont {Gillard},\ and\ \citenamefont
  {Jean}}]{ferriere2007spatial}%
  \BibitemOpen
  \bibfield  {author} {\bibinfo {author} {\bibfnamefont {K.}~\bibnamefont
  {Ferriere}}, \bibinfo {author} {\bibfnamefont {W.}~\bibnamefont {Gillard}},\
  and\ \bibinfo {author} {\bibfnamefont {P.}~\bibnamefont {Jean}},\ }\href
  {https://doi.org/10.1051/0004-6361:20066992} {\bibfield  {journal} {\bibinfo
  {journal} {Astron. Astrophys.}\ }\textbf {\bibinfo {volume} {467}},\ \bibinfo
  {pages} {611} (\bibinfo {year} {2007})}\BibitemShut {NoStop}%
\bibitem [{\citenamefont {Ferriere}(2012)}]{ferriere2012interstellar}%
  \BibitemOpen
  \bibfield  {author} {\bibinfo {author} {\bibfnamefont {K.}~\bibnamefont
  {Ferriere}},\ }\href {https://doi.org/10.1051/0004-6361/201117181} {\bibfield
   {journal} {\bibinfo  {journal} {Astron. Astrophys.}\ }\textbf {\bibinfo
  {volume} {540}},\ \bibinfo {pages} {A50} (\bibinfo {year}
  {2012})}\BibitemShut {NoStop}%
\bibitem [{\citenamefont {Linden}\ \emph {et~al.}(2012)\citenamefont {Linden},
  \citenamefont {Lovegrove},\ and\ \citenamefont
  {Profumo}}]{linden2012morphology}%
  \BibitemOpen
  \bibfield  {author} {\bibinfo {author} {\bibfnamefont {T.}~\bibnamefont
  {Linden}}, \bibinfo {author} {\bibfnamefont {E.}~\bibnamefont {Lovegrove}},\
  and\ \bibinfo {author} {\bibfnamefont {S.}~\bibnamefont {Profumo}},\ }\href
  {https://doi.org/10.1088/0004-637X/753/1/41} {\bibfield  {journal} {\bibinfo
  {journal} {Astrophys. J.}\ }\textbf {\bibinfo {volume} {753}},\ \bibinfo
  {pages} {41} (\bibinfo {year} {2012})}\BibitemShut {NoStop}%
\bibitem [{\citenamefont {Ptuskin}\ \emph {et~al.}(1993)\citenamefont {Ptuskin}
  \emph {et~al.}}]{ptuskin1993diffusion}%
  \BibitemOpen
  \bibfield  {author} {\bibinfo {author} {\bibfnamefont {V.}~\bibnamefont
  {Ptuskin}} \emph {et~al.},\ }\href
  {https://ui.adsabs.harvard.edu/abs/1993A%26A...268..726P/abstract} {\bibfield
   {journal} {\bibinfo  {journal} {Astron. Astrophys.}\ }\textbf {\bibinfo
  {volume} {268}},\ \bibinfo {pages} {726} (\bibinfo {year}
  {1993})}\BibitemShut {NoStop}%
\bibitem [{\citenamefont {Candia}\ \emph {et~al.}(2003)\citenamefont {Candia},
  \citenamefont {Roulet},\ and\ \citenamefont {Epele}}]{candia2003turbulent}%
  \BibitemOpen
  \bibfield  {author} {\bibinfo {author} {\bibfnamefont {J.}~\bibnamefont
  {Candia}}, \bibinfo {author} {\bibfnamefont {E.}~\bibnamefont {Roulet}},\
  and\ \bibinfo {author} {\bibfnamefont {L.~N.}\ \bibnamefont {Epele}},\ }\href
  {https://doi.org/10.1088/1126-6708/2002/12/033} {\bibfield  {journal}
  {\bibinfo  {journal} {J. High Energy Phys.}\ }\textbf {\bibinfo {volume}
  {2002}}\bibinfo  {number} { (12)},\ \bibinfo {pages} {033}}\BibitemShut
  {NoStop}%
\bibitem [{\citenamefont {Evoli}\ \emph {et~al.}(2018)\citenamefont {Evoli},
  \citenamefont {Blasi}, \citenamefont {Morlino},\ and\ \citenamefont
  {Aloisio}}]{evoli2018origin}%
  \BibitemOpen
\bibfield  {number} {  }\bibfield  {author} {\bibinfo {author} {\bibfnamefont
  {C.}~\bibnamefont {Evoli}}, \bibinfo {author} {\bibfnamefont
  {P.}~\bibnamefont {Blasi}}, \bibinfo {author} {\bibfnamefont
  {G.}~\bibnamefont {Morlino}},\ and\ \bibinfo {author} {\bibfnamefont
  {R.}~\bibnamefont {Aloisio}},\ }\href
  {https://doi.org/10.1103/PhysRevLett.121.021102} {\bibfield  {journal}
  {\bibinfo  {journal} {Phys. Rev. Lett.}\ }\textbf {\bibinfo {volume} {121}},\
  \bibinfo {pages} {021102} (\bibinfo {year} {2018})}\BibitemShut {NoStop}%
\bibitem [{\citenamefont {Taillet}\ and\ \citenamefont
  {Maurin}(2003)}]{taillet2003spatial}%
  \BibitemOpen
  \bibfield  {author} {\bibinfo {author} {\bibfnamefont {R.}~\bibnamefont
  {Taillet}}\ and\ \bibinfo {author} {\bibfnamefont {D.}~\bibnamefont
  {Maurin}},\ }\href {https://doi.org/10.1051/0004-6361:20030318} {\bibfield
  {journal} {\bibinfo  {journal} {Astron. Astrophys.}\ }\textbf {\bibinfo
  {volume} {402}},\ \bibinfo {pages} {971} (\bibinfo {year}
  {2003})}\BibitemShut {NoStop}%
\bibitem [{\citenamefont {Beck}(2001)}]{beck2001galactic}%
  \BibitemOpen
  \bibfield  {author} {\bibinfo {author} {\bibfnamefont {R.}~\bibnamefont
  {Beck}},\ }\href {https://doi.org/10.1023/A:1013805401252} {\bibfield
  {journal} {\bibinfo  {journal} {Space Sci. Rev.}\ }\textbf {\bibinfo {volume}
  {99}},\ \bibinfo {pages} {243} (\bibinfo {year} {2001})}\BibitemShut
  {NoStop}%
\bibitem [{\citenamefont {Orlando}\ and\ \citenamefont
  {Strong}(2013)}]{orlando2013galactic}%
  \BibitemOpen
  \bibfield  {author} {\bibinfo {author} {\bibfnamefont {E.}~\bibnamefont
  {Orlando}}\ and\ \bibinfo {author} {\bibfnamefont {A.}~\bibnamefont
  {Strong}},\ }\href {https://doi.org/10.1093/mnras/stt1718} {\bibfield
  {journal} {\bibinfo  {journal} {Mon. Not. R. Astron. Soc.}\ }\textbf
  {\bibinfo {volume} {436}},\ \bibinfo {pages} {2127} (\bibinfo {year}
  {2013})}\BibitemShut {NoStop}%
\bibitem [{\citenamefont {Asakimori}\ \emph {et~al.}(1998)\citenamefont
  {Asakimori} \emph {et~al.}}]{asakimori1998cosmic}%
  \BibitemOpen
  \bibfield  {author} {\bibinfo {author} {\bibfnamefont {K.}~\bibnamefont
  {Asakimori}} \emph {et~al.},\ }\href {https://doi.org/10.1086/305882}
  {\bibfield  {journal} {\bibinfo  {journal} {Astrophys. J.}\ }\textbf
  {\bibinfo {volume} {502}},\ \bibinfo {pages} {278} (\bibinfo {year}
  {1998})}\BibitemShut {NoStop}%
\bibitem [{\citenamefont {Apanasenko}\ \emph {et~al.}(2001)\citenamefont
  {Apanasenko} \emph {et~al.}}]{balloon2001composition}%
  \BibitemOpen
  \bibfield  {author} {\bibinfo {author} {\bibfnamefont {A.}~\bibnamefont
  {Apanasenko}} \emph {et~al.} (\bibinfo {collaboration} {RUNJOB
  Collaboration}),\ }\href {https://doi.org/10.1016/S0927-6505(00)00163-8}
  {\bibfield  {journal} {\bibinfo  {journal} {Astropart. Phys.}\ }\textbf
  {\bibinfo {volume} {16}},\ \bibinfo {pages} {13} (\bibinfo {year}
  {2001})}\BibitemShut {NoStop}%
\bibitem [{\citenamefont {Finger}(2011)}]{finger2011reconstruction}%
  \BibitemOpen
  \bibfield  {author} {\bibinfo {author} {\bibfnamefont {M.~R.}\ \bibnamefont
  {Finger}},\ }\href@noop {} {Ph.D. thesis},\ \bibinfo  {school} {Karlsruher
  Institut f{\"u}r Technologie (KIT)} (\bibinfo {year} {2011})\BibitemShut
  {NoStop}%
\bibitem [{\citenamefont {Akiyama}\ \emph {et~al.}(2021)\citenamefont {Akiyama}
  \emph {et~al.}}]{akiyama2021first}%
  \BibitemOpen
  \bibfield  {author} {\bibinfo {author} {\bibfnamefont {K.}~\bibnamefont
  {Akiyama}} \emph {et~al.} (\bibinfo {collaboration} {Event Horizon Telescope
  Collaboration}),\ }\href {https://doi.org/10.3847/2041-8213/abe4de}
  {\bibfield  {journal} {\bibinfo  {journal} {Astrophys. J. Lett.}\ }\textbf
  {\bibinfo {volume} {910}},\ \bibinfo {pages} {L13} (\bibinfo {year}
  {2021})}\BibitemShut {NoStop}%
\bibitem [{\citenamefont {Mo{\'s}cibrodzka}\ \emph {et~al.}(2016)\citenamefont
  {Mo{\'s}cibrodzka}, \citenamefont {Falcke},\ and\ \citenamefont
  {Shiokawa}}]{moscibrodzka2016general}%
  \BibitemOpen
  \bibfield  {author} {\bibinfo {author} {\bibfnamefont {M.}~\bibnamefont
  {Mo{\'s}cibrodzka}}, \bibinfo {author} {\bibfnamefont {H.}~\bibnamefont
  {Falcke}},\ and\ \bibinfo {author} {\bibfnamefont {H.}~\bibnamefont
  {Shiokawa}},\ }\href {https://doi.org/10.1051/0004-6361/201526630} {\bibfield
   {journal} {\bibinfo  {journal} {Astron. Astrophys.}\ }\textbf {\bibinfo
  {volume} {586}},\ \bibinfo {pages} {A38} (\bibinfo {year}
  {2016})}\BibitemShut {NoStop}%
\end{thebibliography}%

\end{document}